\documentclass[aps,prx,twocolumn,superscriptaddress,floatfix]{revtex4-2}
\usepackage{adjustbox}
\usepackage{graphicx}
\usepackage{dcolumn}
\usepackage{bm}
\usepackage{xcolor}
\usepackage[inkscapearea=page]{svg}
\usepackage{soul}
\usepackage[T1]{fontenc}
\usepackage{amstext}
\usepackage{amsmath}
\usepackage{amsfonts}
\usepackage{amssymb}
\usepackage{float}
\usepackage{algorithm}
\usepackage{algpseudocode}
\usepackage{appendix}
\usepackage{multirow}
\usepackage{physics}
\usepackage[colorlinks=true,linkcolor=blue,
urlcolor=black,anchorcolor=blue,
citecolor=blue]{hyperref}
\usepackage[percent]{overpic}
\usepackage{enumerate}
\usepackage{subfigure}
\usepackage{upgreek}
\begin{document}
\preprint{APS/123-QED}

\title{Photon-Number Conserved Universal Quantum Logic Employing Continuous-Time Quantum Walk on Dual-Rail Qubit Arrays}
\author{Hao-Yu Guan}
\thanks{These authors contributed equally}
\affiliation{Shenzhen Institute for Quantum Science and Engineering, Southern University of Science and Technology, Shenzhen, Guangdong 518055, China}
\author{Yifei Li}
\thanks{These authors contributed equally}
\affiliation{Cuiying Honors College, Lanzhou University, Lanzhou 730000, China}
\author{Xiu-Hao Deng}
\email{dengxh@sustech.edu.cn}
\affiliation{Shenzhen Institute for Quantum Science and Engineering, Southern University of Science and Technology, Shenzhen, Guangdong 518055, China}
\affiliation{International Quantum Academy, Shenzhen, Guangdong 518000, China}
\date{\today}

\begin{abstract}
We demonstrate a synergy between dual-rail qubit encoding and continuous-time quantum walks (CTQW) to realize universal quantum logic in superconducting circuits. Utilizing the photon-number-conserving dynamics of CTQW on dual-rail transmons, which systematically transform leakage and relaxation into erasure events, our architecture facilitates the suppression of population leakage and the implementation of high-fidelity quantum gates. We construct single-, two-, and three-qubit operations that preserve dual-rail encoding, facilitated by tunable coupler strengths compatible with current superconducting qubit platforms. Numerical simulations confirm robust behavior against dephasing, relaxation, and imperfections in coupling, underscoring the erasure-friendly nature of the system. This hardware-efficient scheme thus provides a practical pathway to early fault-tolerant quantum computation.
\end{abstract}
\maketitle

\section{Introduction}
In quantum information processing, high-fidelity control and error correction are vital, as superconducting architectures often suffer from leakage out of the computational subspace or relaxation to lower-energy states. A promising path to address these issues is converting leakage or relaxation events into erasure errors, which explicitly flag themselves for correction~\cite{kubica2023erasure, shim2016semiconductor}. Recent studies on dual-rail transmons encode information in a single photon-number excitation distributed across two resonantly coupled transmons, enabling efficient leakage detection and improving error-correction thresholds~\cite{teoh2023dual,campbell2020universal, levine2024demonstrating,chou2024superconducting, koottandavida2024erasure,weiss2024quantum,wong2019isolated,chawla2023multi}.

In parallel, continuous-time quantum walks (CTQW) provide a powerful framework that preserves the total number of excitations, making them a natural match for dual-rail encoding. By confining the walker—a single excitation—to a well-defined subspace, CTQW safeguards against leakage while enabling potential quantum speedups in hitting and mixing processes~\cite{farhi1998quantum, childs2002example, childs2003exponential, childs2004quantum}. These features have spurred experimental demonstrations across various integrated qubit systems, such as superconducting qubits~\cite{peruzzo2010quantum, yan2019strongly, gong2021quantum}. Beyond single-particle dynamics, interactions between multiple walkers introduce richer correlated behavior~\cite{lahini2012quantum, siloi2017noisy} and can support robust quantum search~\cite{lewis2021optimal, xing2024quantum,wong2016spatial,wong2015grover} and error correction~\cite{wang2022multiparticle}. From a theoretical standpoint, multi-walkers CTQW can construct universal quantum logic~\cite{childs2013universal,underwood2012bose}, supported by proposals demonstrating high-fidelity CPhase gates~\cite{e2024two} or leveraging two internal states of the walker on a directed graph~\cite{asaka2023two}. A CNOT gate has been constructed for both non-interacting bosons—realized by photons in waveguide lattices—and for interacting bosons using ultra-cold atoms~\cite{lahini2018quantum}. All these advancements highlight the potential of CTQW in universal quantum computing.

By integrating the dual-rail transmon encoding with CTQW, it is possible to harness the advantages of both methodologies: robust error management via erasure conversion and the preservation of photon-number properties inherent to CTQW. This integration enhances the efficiency of erasure error correction in quantum computing architectures built upon this framework. In this study, we propose a foundational framework to implement universal quantum logic based on the operation of correlated CTQW within the structure of dual-rail encoded qubit arrays. We present explicit constructions of single- and two-qubit gates, such as the controlled-Z and iSWAP gates, alongside three-qubit gates, maintaining the dual-rail encoding by the end of the quantum evolution. Our analysis encompasses both transverse and longitudinal connections within a superconducting qubit array, deriving parameter regimes that align with current experimental methodologies utilizing tunable transmon couplers. Additionally, we investigate the behavior of these gate constructions under realistic noise conditions, including dephasing, relaxation, and imperfections in coupler strengths or detunings. Our results suggest that the intrinsic properties of dual-rail transmons, in conjunction with the excitation-conserving attributes of CTQW, render this approach not only feasible but potentially robust against prevalent hardware imperfections. This comprehensive perspective holds promise for advancing early efforts in fault-tolerant quantum computation by providing both practical means for gate synthesis and a stable encoding architecture for superconducting circuits.

\section{dual-rail transmon array}\label{sec: physical qubit}

\begin{figure}[t]
	\centering
	\includegraphics[width=1\columnwidth]{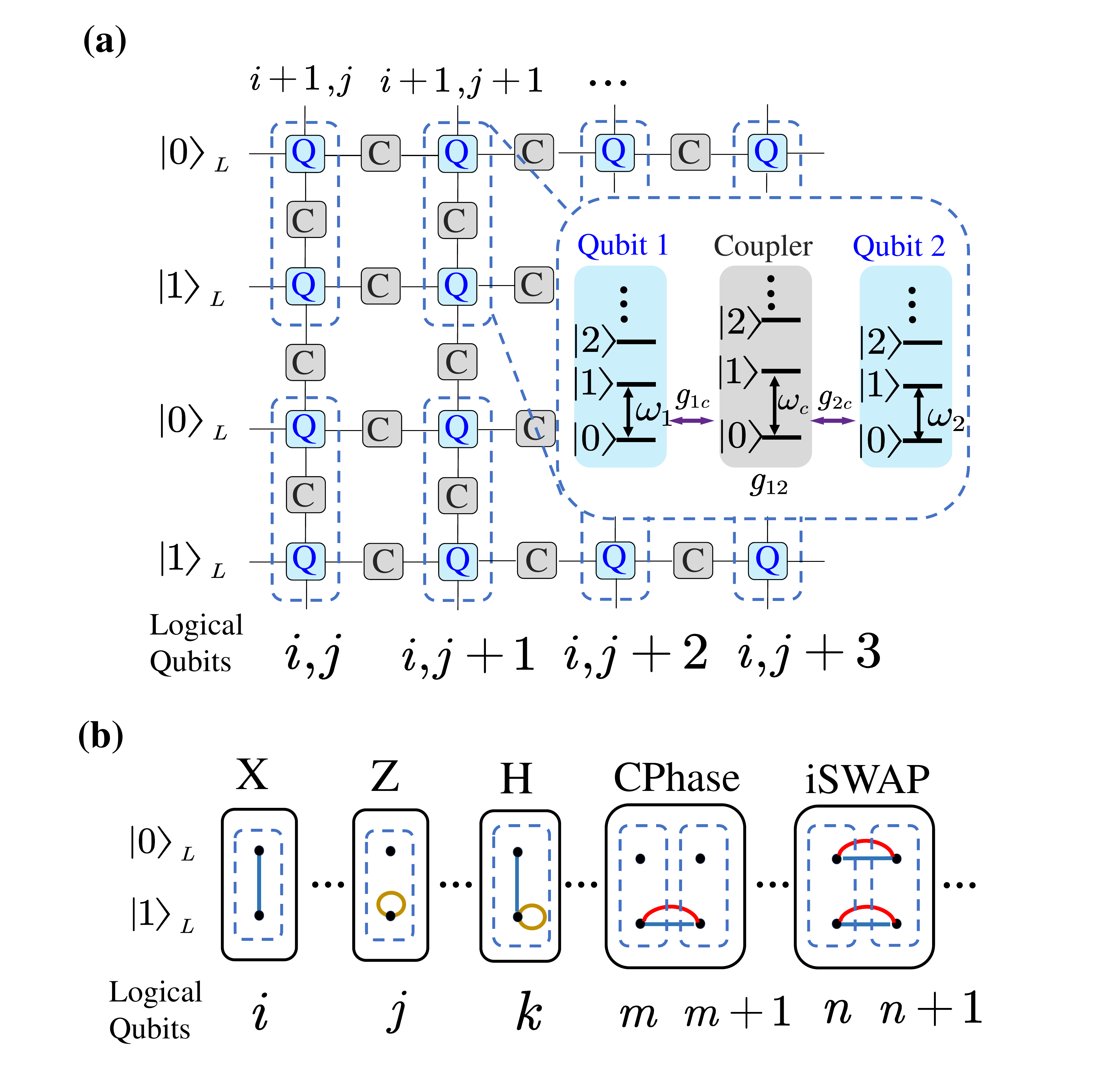}
	\caption{\textbf{Illustration of the dual-rail encoded qubit array and logical operations.} 
(a) A building block of a $4\times 2$ logical qubits array. Each logical qubit (dashed boxes) consists of two transmon (blue boxes) and one coupler (gray boxes). The coupling scheme is illustrated in the inset dashed box.
(b) Quantum walk on various graphs represent different logical operations on the 1-D logical qubit array, with corresponding gates marked above the solid boxes. Each vertex corresponds to a transmon, while a pair of transmons in the dashed box indicates an encoded logical qubit. Blue, orange, and red edges signify translational, loop-back, and collision walks, respectively.}
	\label{Fig: Illustrating}
\end{figure}

\subsection{Physical Qubit with Tunable Couplers}
We begin with a prototype 2-D superconducting circuit consisting of physical qubits and tunable couplers, as depicted in Fig.~\ref{Fig: Illustrating}(a). To optimize coherence and facilitate experimental control, each physical qubit and coupler is implemented using a tunable transmon~\cite{koch2007charge, hutchings2017tunable}. These transmons can be accurately approximated as Duffing oscillators, a common model for anharmonic multilevel qubit systems, which also includes capacitively shunted flux qubits~\cite{steffen2010high, yan2016flux}. In this setup, the quantum bits are encoded in the first two energy levels of the transmons. Nevertheless, we shall also treat the third level as a resource for phase accumulations, which will be discussed when constructing the CPhase gate.

A simple and versatile tunable coupling scheme, illustrated in the inset of Fig.~\ref{Fig: Illustrating}(a), was proposed and experimentally realized~\cite{yan2018tunable, sung2021realization}. This scheme involves a pairwise interacting three-body quantum system, where the leftmost and rightmost anharmonic oscillators function as the physical qubits while the middle oscillator serves as the coupler. These two qubits and the coupler interact through exchange-type interactions, with coupling strengths $g_{1c}$, $g_{2c}$, and $g_{12}$, respectively. The Hamiltonian describing this three-body system is given by:
\begin{equation}
\hat{H} = \sum_i \left( \omega_i \hat{b}_i^\dagger \hat{b}_i + \frac{\beta_i}{2}  \hat{b}_i^\dagger \hat{b}_i^\dagger \hat{b}_i  \hat{b}_i  \right) 
+ \sum_{i<j} g_{ij} \left(  \hat{b}_i -  \hat{b}_i^\dagger \right) \left(  \hat{b}_j - \hat{b}_j^\dagger \right),
\end{equation}
where $i, j \in \{1, 2, c\}$, $\omega_i$ is the oscillator frequency, and $\beta_i$ denotes the anharmonicity of the oscillator. The operators $\hat{b}_i^\dagger$ and $\hat{b}_i$ are the creation and annihilation operators, respectively, defined in the eigenbasis of the corresponding oscillator. We assume that the coupler-qubit interactions are significantly stronger than the direct qubit-qubit interaction, i.e., $g_{1c}, g_{2c} \gg g_{12}$, which corresponds to the practical parameter regime for tunable couplers~\cite{yan2018tunable}.
 
The three-body system serves as a unit cell for encoding a logical qubit, defined in the system's first excitation subspace. Before proceeding, we derive an effective two-qubit system by applying a block diagonalization transformation to decouple the coupler. We assume that both qubits are negatively detuned from the coupler, $\Delta_i \equiv \omega_i - \omega_c < 0$, and that the coupling is dispersive, $g_{ic} \ll \left|\Delta_i\right| \ (i=1,2)$. Additionally, we assume that the coupler mode always remains in its ground state, with its excitations not participating in the system dynamics.

As shown in Ref.~\cite{yan2018tunable}, the destructive interference between the coupler-induced effective and direct qubit-qubit couplings enables the net coupling to be turned on or off by tuning the coupler frequency $\omega_c$. A small ZZ interaction can be perturbatively derived up to the fourth order in $g_i/\Delta_i$ using Bloch's theory~\cite{takayanagi2016effective}. Furthermore, exact results of the effective Hamiltonian can be determined from the least action principle~\cite{cederbaum1989block}.

To a good approximation, the physics of the two-transmon building block is well captured by the extended Bose-Hubbard model (EBHM). In the qubit regime, this reduces to a Heisenberg XXZ spin model. The effective two-body Hamiltonian is expressed as:
\begin{equation}
\begin{aligned}
    \hat{H}^{\rm eff} = & \sum_{i=1,2} \left( \tilde{\omega}_i \hat{n}_i + \frac{\beta_i}{2} \hat{b}_i^\dagger \hat{b}_i^\dagger \hat{b}_i \hat{b}_i \right) 
    + \tilde{g} \left( \hat{b}_1^\dagger \hat{b}_2 + \hat{b}_1 \hat{b}_2^\dagger \right) \\
    & + V \hat{n}_1 \hat{n}_2,
\end{aligned}
\end{equation}
where $\hat{n}_i =  \hat{b}_i^\dagger  \hat{b}_i$ is the excitation number operator, $\tilde{\omega}_i = \omega_i + g_i^2 / \Delta_i$ is the Lamb-shifted qubit frequency, $\tilde{g} = g_{1c} g_{2c} / \Delta + g_{12}$ is the effective coupling strength, $1 / \Delta = \left( 1 / \Delta_1 + 1 / \Delta_2 \right) / 2$, and 
\begin{equation}\label{eq: ZZ_RW}
	\begin{aligned}
		V=&4g_1g_2g_{12}\left[\frac{1}{\Delta_{1}^2}-\frac{1}{\Delta_{1}(\Delta_{1}+\beta_1)}\right]\\
		+&4g_1^2g_2^2\left[\frac{2}{\Delta_1^2(2\Delta_{1}-\beta_c)}-\frac{1}{\Delta_{1}^2(\Delta_{1}+\beta_1)}\right]
	\end{aligned}
\end{equation}
represents the ZZ interaction strength. Typically, the ZZ interaction is an order of magnitude smaller than the coupling strength $\tilde{g}$.
This formalism can be readily extended to model a one-dimensional qubit chain or a two-dimensional qubit array.

\subsection{Continuous-time Quantum Walks on the Extended Bose-Hubbard Model}

The EBHM serves as a nontrivial model for studying quantum walks. When there is a single walk on graphs, there is no additional interaction energy; when multiple walkers are on the same graphs, they interact with each other and exhibit correlated quantum walks~\cite{lahini2018quantum}. These quantum walks go beyond simple hopping, incorporating local phase shifts, on-site interactions, and nearest-neighbor correlations associated with EBHM. This section introduces the EBHM and its graph-based interpretation as a platform for describing such nontrivial CTQWs.

The extended Bose--Hubbard model (EBHM) incorporates additional interactions beyond the standard Bose--Hubbard framework, enabling the study of diverse quantum phenomena~\cite{dutta2015non}. It has been experimentally realized in ultracold atoms in optical lattices~\cite{baier2016extended}, Rydberg atom arrays~\cite{browaeys2020many,ebadi2022quantum}, superconducting circuits~\cite{kounalakis2018tuneable,levine2024demonstrating}, and dipolar excitons~\cite{lagoin2022extended}. The EBHM Hamiltonian takes the form
\begin{equation}
\hat{H} 
= -J \sum_{\langle i, j \rangle} \hat{c}_i^\dagger \hat{c}_j 
+ \frac{U}{2} \sum_i \hat{n}_i \bigl(\hat{n}_i - 1\bigr) 
- \mu \sum_i \hat{n}_i 
+ V \sum_i \hat{n}_i \hat{n}_{i+1},
\end{equation}
where $\hat{c}_i^{(\dagger)}$ annihilates (creates) a boson at site $i$, $\hat{n}_i \equiv \hat{c}_i^\dagger \hat{c}_i$ is the number operator, and the parameters $J$, $U$, $\mu$, and $V$ denote the tunneling amplitude, on-site interaction strength, chemical potential, and nearest-neighbor interaction strength, respectively.

In this context, a graph \(G = (V_G, E)\) is defined, where \(V_G\) represents the lattice sites (vertices) and \(E\) denotes the edges that correspond to specific Hamiltonian terms. To visually represent the physical processes described by the Hamiltonian, we employ the following color coding:
\begin{itemize}
    \item Blue Edges: Correspond to the tunneling term (\(-J \sum_{\langle i, j \rangle} \hat{c}_i^\dagger \hat{c}_j\)) and describe particle hopping between neighboring sites.
    \item Orange Edges: Represent the chemical potential term (\(-\mu \sum_i \hat{n}_i\)) and capture local phase accumulation or energy shifts.
    \item Red Edges: Represent the nearest-neighbor interaction term (\(V \sum_i \hat{n}_i \hat{n}_{i+1}\)), modeling interactions between particles at adjacent sites.
\end{itemize}
To align with superconducting circuit implementations, we will henceforth refer to these terms as the coupling, frequency shift (detuning), and ZZ interaction terms, respectively.

In the language of CTQWs, these edges describe distinct types of quantum walks, as depicted in Fig. \ref{Fig: Illustrating}(b):
\begin{itemize}
    \item \textit{Translational Walk}: (Blue Edges) A walker hops from one site to another.
    \item \textit{Loop-Back Walk}: (Orange Edges) A walker remains confined to a single site and accumulates phase, analogous to self-loops in a graph.
    \item \textit{Collision Walk}: (Blue + Red Edges) Neighboring walkers exchange sites with interaction, resulting in correlated quantum walk. 
    \item \textit{Collision}: (Red Edges) Neighboring walkers interact, resulting in pure interaction. 
\end{itemize}

Each graph $G$ can be transformed into a dual graph, which encodes the structure of couplings and interactions among the Fock states governed by the Hamiltonian. This dual graph provides an intuitive graphical representation of the transitions facilitated by the Hamiltonian terms, illustrating how quantum states interact and evolve.


This graph-based approach opens pathways for studying many-body quantum dynamics and correlations in a controlled experimental setting.

\subsection{Dual-Rail Encoding}
\label{subsec:dual-rail}

In this subsection, we describe a dual-rail encoding scheme in the EBHM, where single microwave-photonic excitations (``walkers'') propagate on a graph $G$. This encoding method exploits the spatial delocalization of the walker to represent qubits across pairs of adjacent transmons. We introduce the computational and complementary subspaces, discuss the total Hilbert space dimension, and explain the default system configuration used for qubit initialization.

\medskip

As depicted in Fig.~\ref{Fig: Illustrating}, a single walker occupying two adjacent transmons constitutes a logical qubit. The logical states \(\ket{0}_L\) and \(\ket{1}_L\) are defined by the excitation being localized in the upper or lower site, respectively. In a superposition, the walker delocalizes across these two sites, enabling qubit encoding without additional degrees of freedom. 

Consider a Hamiltonian graph \(G\) with \(2 \times n\) vertices (\(2n\) transmons). Placing \(n\) independent walkers on these vertices encodes a chain of \(n\) qubits, while the total dimension of the full Hilbert space is
\begin{equation}
\mathcal{D} = \binom{3n-1}{n} = \frac{(3n-1)!}{n!(2n-1)!} \approx 2.6^n,
\end{equation}
which is substantially larger than the \(2^n\)-dimensional computational subspace \(\mathcal{H}_{\mathrm{C}}\). We decompose the entire Hilbert space as 
\begin{equation}
\mathcal{H} = \mathcal{H}_{\mathrm{C}} \oplus \mathcal{H}_{\perp},
\end{equation}
where \(\mathcal{H}_{\perp}\) denotes the complementary subspace that does not encode qubit information.

Label the transmon vertices as \(v_{i,x} \in V_G\), where \(i \in \{1,\ldots,n\}\) indexes columns and \(x \in \{0,1\}\) indexes rows. The vacuum state of the entire system is
\begin{equation}
\ket{\mathrm{vac}} = \ket{0_{1,0} 0_{1,1};\cdots;0_{n,0} 0_{n,1}},
\end{equation}
where semicolons separate different logical columns. A walker created at \(v_{i,x}\) is given by \(\hat{c}_{i,x}^\dagger\ket{\mathrm{vac}}\). Each two-site column \((i,0)\) and \((i,1)\) comprises a logical qubit, with the basis states for $\mathcal{H}_{\mathrm{C}} $
\begin{equation}
\begin{aligned}
    \ket{0_i}_L &\equiv \ket{1_{i,0} 0_{i,1}},\\
    \ket{1_i}_L &\equiv \ket{0_{i,0} 1_{i,1}},
\end{aligned}
\end{equation}
so that an \(n\)-walker state serves as a valid computational state if and only if each column contains exactly one excitation:
\begin{equation}
\sum_{x=0}^1 \bigl|\langle \psi \bigr| \hat{c}_{i,x}^\dagger \hat{c}_{i,x} \bigl|\psi\rangle \bigr|^2 = 1, 
\quad i \in \{1,\ldots,n\}.
\end{equation}

Any state \(\ket{\Psi}\) with support in \(\mathcal{H}_{\perp}\) does not encode valid qubit information. Although a physical Hamiltonian may couple \(\mathcal{H}_{\mathrm{C}}\) to \(\mathcal{H}_{\perp}\) and allow the walker to transiently explore \(\mathcal{H}\), a well-defined gate time \(T\) ensures that the resulting unitary remains block-diagonal:
\begin{equation}
U = U_{\mathrm{C}} \oplus U_{\perp}.
\end{equation}
Hence, any state initially in \(\mathcal{H}_{\mathrm{C}}\) returns to the computational subspace at the end of the evolution---despite possible temporary transitions into \(\mathcal{H}\). This mechanism leverages the larger Hilbert space to generate entanglement, aided by bosonic indistinguishability~\cite{benatti2020entanglement}.

Unless stated otherwise, we assume coupling strength \(J=0\), frequency shift \(\mu=0\), and ZZ interaction strength \(V=0\). In this default configuration, the system Hamiltonian is
\begin{equation}
\hat{H}_0 
= \frac{U}{2}\sum_{i=1}^n \sum_{x=0}^1
\hat{n}_{i,x}\bigl(\hat{n}_{i,x}-1\bigr),
\end{equation}
resulting in a disconnected graph of \(2n\) vertices with no edges. Initializing the system in \(\ket{\psi}\in \mathcal{H}_{\mathrm{C}}\) yields a zero-energy eigenstate of \(\hat{H}_0\), ensuring no dynamics occur in the absence of applied operations.

\section{Photon-number conserved universal gates}

This section delineates the implementation of universal sets of single-qubit gates and assorted two-qubit gates, including the CPhase, iSWAP, and CCPhase gates within transverse connections, as well as the CPhase gate in longitudinal connections. The realization of these gates involves activating one or two couplings in the system, thereby transiently linking the computational subspace \( \mathcal{H}_{\rm C} \) with its orthogonal complement \( \mathcal{H}_{\perp} \). A pivotal challenge is the precise determination of the gate duration \( T \) and the selection of an appropriate Hamiltonian, ensuring that by the conclusion of the operation, the state evolution remains confined to the computational subspace. As depicted in Fig.~\ref{Fig: Illustrating}(a), the dual-rail encoded qubits positioned in a two-dimensional array showcase anisotropic interconnections. Transverse connections accommodate two coupling channels, whereas longitudinal connections possess a solitary coupling channel. This anisotropy necessitates distinct strategies for gate implementation, customized to the specific type of connection. To furnish a comprehensive overview, a succinct introduction to single-qubit gate implementation precedes the discussion of two-qubit gate constructions.

\subsection{Single-Qubit Gates}

Here, we demonstrate the implementation of single-qubit gates \( X \), \( Z \) with arbitrary rotation angles, in addition to the Hadamard gate. The associated quantum walk graphs are illustrated in Fig.~\ref{Fig: Illustrating}(b).


The Hamiltonian for the \( X \) gate applied to the \( i \)-th logical qubit is:
\begin{equation}
\hat{H}_{X,i} = -J_{X} \left( \hat{c}_{i,1}^{\dagger} \hat{c}_{i,0} + \text{H.c.} \right) + \mu_X(\hat{n}_{i,0}+\hat{n}_{i,1}) + \hat{H}_0,
\end{equation}
where \( J_X \) is the coupling strength and \( \mu_X \) is the frequency shift of the two physical qubits. When restricted to the two computational states \( \ket{0_i}_L \) and \( \ket{1_i}_L \), the Hamiltonian simplifies to \( -J_X \hat{\sigma}_x \), where \( \hat{\sigma}_x \) is the Pauli \( X \) operator. The corresponding gate time for the \( X \)-gate is:
\begin{equation}
T_X = \frac{\pi}{2J_X}.
\end{equation}
The global phase induced by the coupling term can be canceled by setting \( \mu_X = J_X \).

For the \( Z \) gate on the \( j \)-th logical qubit, the Hamiltonian becomes:
\begin{equation}
\hat{H}_{Z,j} = - \mu_{Z} \hat{n}_{j,1} + \hat{H}_0,
\end{equation}
where \( \mu_{Z} \) represents the detuning applied to the site \( (j,1) \). The Hamiltonian in the computational subspace is:
\begin{equation}
\hat{H}_{Z,j} = \begin{pmatrix}
0 & 0 \\
0 & -\mu_{Z}
\end{pmatrix}.
\end{equation}
This Hamiltonian induces a phase shift on \( \ket{1_j}_L \) at a rate \( -\mu_Z \). The gate time for the \( Z \)-gate is:
\begin{equation}
T_Z = \frac{\pi}{\mu_Z}.
\end{equation}


Since both \( X \) and \( Z \) gates do not couple states in \( \mathcal{H_{\text{C}}} \) to those in \( \mathcal{H_{\perp}} \), they can be easily extended to arbitrary \( X \)-rotations \( \hat{R}_X(\theta) = e^{-i\hat{\sigma}_x \theta} \), with gate time:
\begin{equation}
T_X(\theta) = \frac{2\pi - \theta}{J_X},
\end{equation}
and phase gates
\begin{equation}
P(\theta) = \begin{pmatrix}
1 & 0 \\
0 & e^{i\theta}
\end{pmatrix},
\end{equation}
with gate time:
\begin{equation}
T_P(\theta) = \frac{\theta}{\mu_Z}.
\end{equation}
Arbitrary single-qubit gates can thus be implemented by decomposing rotations on the Bloch sphere using Euler angles \cite{nielsen2001quantum}.


The Hadamard gate can be directly implemented on the \( k \)-th qubit using the Hamiltonian:
\begin{equation}
\hat{H}_{H,k} = \mu_{H} \hat{n}_{k,1} - J_{H} \left( \hat{c}_{k,0}^{\dagger} \hat{c}_{k,1} + \text{H.c.} \right) + \hat{H}_0.
\end{equation}
For \( \mu_H / J_H = 2 \), this Hamiltonian results in a Hadamard gate with a gate time:
\begin{equation}
T_H = \frac{\pi}{2\sqrt{2} J_H}.
\end{equation}
This is more efficient than applying three separate decomposed single-qubit gates. The overall phase can be eliminated by adding a detuning term to both sites, with \( \mu / J_H = \sqrt{2} - 1 \).

\subsection{CPhase Gate for Transverse Connections}

This subsection discusses the construction of controlled-phase (CPhase) gate in transverse connections, leaving the longitudinal case for the next subsection. In particular, we show how a weak ZZ interaction can be utilized to implement a controlled-Z (CZ) gate in a superconducting circuit. 

\begin{figure}[t]
	\centering
	\includegraphics[width=\columnwidth]{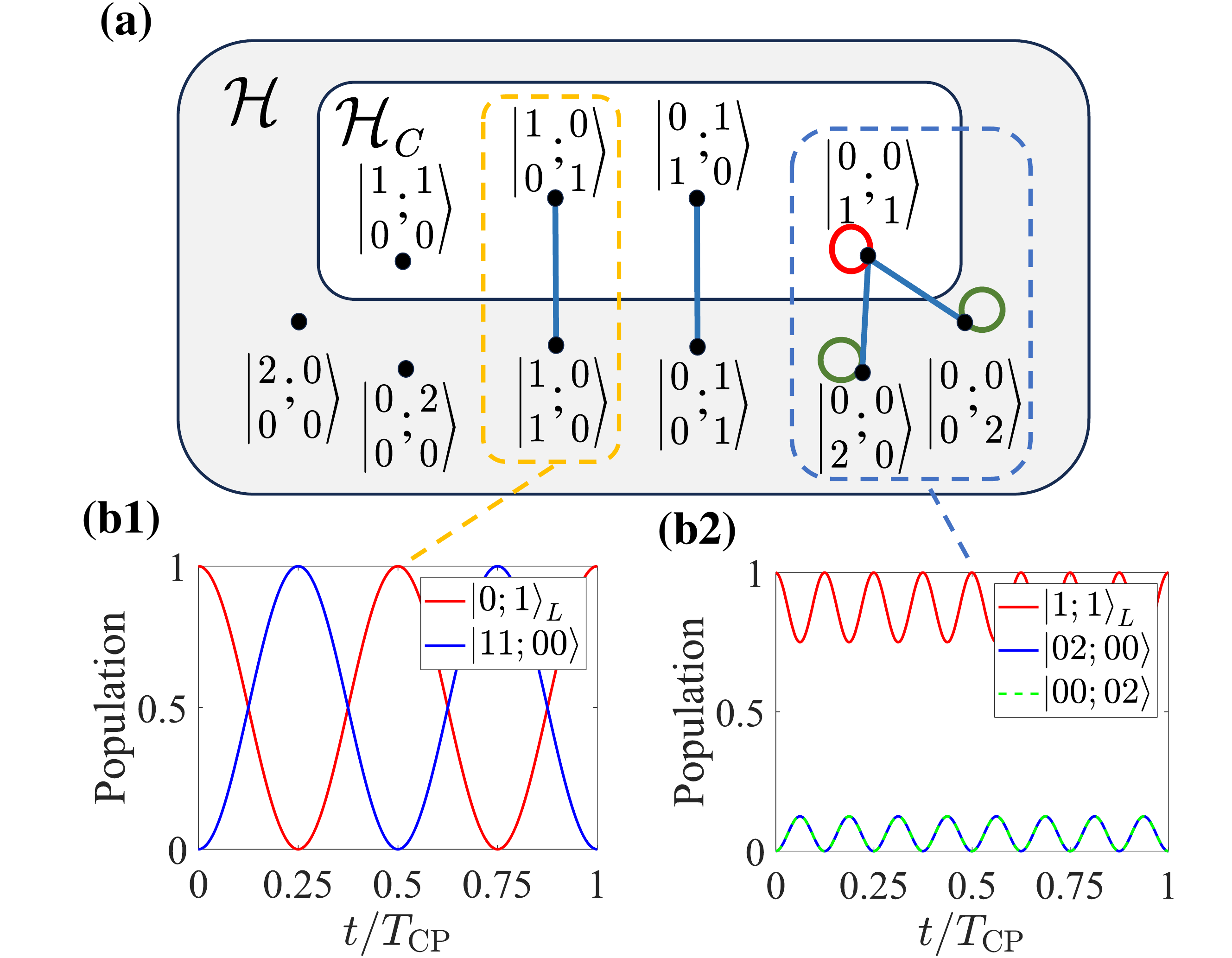}
	\caption{
	\textbf{Dual graph and evolutions of the CPhase gate on dual-rail encoded qubits.}
	(a) The graph describing the couplings and interactions among Fock states under the Hamiltonian $\hat{H}_{\text{CP}}$. The blue edges represent allowed couplings, the red self-loop denotes the ZZ interaction, and the green self-loops represent on-site interactions. 
	(b) Population evolution of states within the dashed boxes in (a). The $x$-axis represents time in units of the gate time $T_{\rm CP}$, and the $y$-axis shows the population of each state. The computational basis states return to their original population by the end of the gate time.
	}
	\label{fig: CZgate}
\end{figure}

We begin by considering a $2 \times 2$ physical qubit block in a transverse connection for logical qubits, as shown in Fig.~\ref{Fig: Illustrating}(b). In this subsection, we label the four vertices as $v_{i,0}$, $v_{i+1,0}$, $v_{i,1}$, and $v_{i+1,1}$. There are two available coupling channels between adjacent qubits: one between $v_{i,0}$ and $v_{i+1,0}$, and another between $v_{i,1}$ and $v_{i+1,1}$. These couplings are distinct from the longitudinal connection, which supports only a single coupling.

For the CPhase gate, we consider a system of two bosonic walkers on four lattice sites, resulting in a 10-dimensional Fock space $\mathcal{H}$. As shown in Fig.~\ref{fig: CZgate}(a), the computational subspace $\mathcal{H}_{\rm C}$ is spanned by the following four Fock states:
\begin{equation}
\begin{aligned}
    \hat{c}_{i, 0}^{\dagger} \hat{c}_{i+1,0}^{\dagger}|00;00\rangle &= |10;10\rangle \leftrightarrow \ket{0;0}_L, \\
    \hat{c}_{i, 0}^{\dagger} \hat{c}_{i+1,1}^{\dagger}|00;00\rangle &= |10;01\rangle \leftrightarrow \ket{0;1}_L, \\
    \hat{c}_{i, 1}^{\dagger} \hat{c}_{i+1,0}^{\dagger}|00;00\rangle &= |01;10\rangle \leftrightarrow \ket{1;0}_L, \\
    \hat{c}_{i, 1}^{\dagger} \hat{c}_{i+1,1}^{\dagger}|00;00\rangle &= |01;01\rangle \leftrightarrow \ket{1;1}_L.
\end{aligned}
\end{equation}
The remaining six states span the orthogonal subspace $\mathcal{H}_{\perp}$.

The CPhase gate is implemented by activating the coupling between the sites $(i,1)$ and $(i+1,1)$ along with the ZZ interaction. The system Hamiltonian is given by:
\begin{equation}
\hat{H}_{\text{CP}} = -J_{\rm CP}(\hat{c}_{i,1}^{\dagger} \hat{c}_{i+1,1} + \text{H.c.}) + V_{\rm CP} \hat{n}_{i,1} \hat{n}_{i+1,1} + \hat{H}_0,
\end{equation}
where $\hat{H}_0$ is the default Hamiltonian of the system.

The evolution of the system is governed by the unitary operator $\hat{U}_{\rm CP}(t) = \exp(-i\hat{H}_{\rm CP} t)$. Since the computational state $\ket{0;0}_L$ is decoupled from all other states, its evolution is trivial:
\begin{equation}
\hat{U}_{\rm CP}(t) \ket{10;10} = \ket{10;10}.
\end{equation}

For the second computational state $\ket{0;1}_L$, we find that it couples to the non-computational state $\ket{11;00}$. The evolution is given by:
\begin{equation}
\hat{U}_{\rm CP}(t) \ket{10;01} = \cos(J_{\rm CP} t) \ket{10;01} + i \sin(J_{\rm CP} t) \ket{11;00}.
\end{equation}
To keep the state within the computational subspace, the evolution must be constrained by a specific gate time $T_{\rm CP}$, which we define as:
\begin{equation}\label{eq: CP time}
    T_{\rm CP} = \frac{2\pi}{|J_{\rm CP}|}.
\end{equation}

Similar dynamics hold for the third computational state $\ket{1;0}_L$. For the fourth state $\ket{1;1}_L$, additional complexities arise as it couples to the non-computational states $\ket{02;00}$ and $\ket{00;02}$. These states will accumulate additional phases, either due to the on-site interaction or the ZZ interaction.

Thus, the evolution operator $\hat{U}_{\rm CP}(t)$ can be decomposed into block diagonal form, with the Hamiltonian being block-diagonal within four invariant subspaces. We define these subspaces as:
\begin{equation}
    \label{eq: four subspaces}
\begin{aligned}
    \mathcal{H}_1 &= \left\{\ket{0;0}_L\right\}, \\
    \mathcal{H}_2 &= \text{span}\left\{\ket{0;1}_L,\ket{11;00}\right\}, \\
    \mathcal{H}_3 &= \text{span}\left\{\ket{1;0}_L,\ket{00;11}\right\}, \\
    \mathcal{H}_4 &= \text{span}\left\{\ket{1;1}_L,\ket{02;00},\ket{00;02}\right\}.
\end{aligned}
\end{equation}

The Hamiltonian within the fourth subspace $\mathcal{H}_4$ is represented by the matrix:
\begin{equation}\label{eq: 3_d_Hamiltonian}
    \hat{H}_4 = \begin{pmatrix}
V & -\sqrt{2}J & -\sqrt{2}J \\
-\sqrt{2}J & U & 0 \\
-\sqrt{2}J & 0 & U
\end{pmatrix}.
\end{equation}
We derive the eigenstates and eigenvalues of $\hat{H}_4$ as follows:
\begin{equation}
\begin{aligned}
    \ket{\phi_1} &= \frac{1}{\sqrt{2}}\begin{pmatrix} 0 \\ -1 \\ 1 \end{pmatrix}, \\
    \ket{\phi_2} &= \alpha_2 \begin{pmatrix} \frac{\sqrt{2}(U-V+X)}{4J} \\ 1 \\ 1 \end{pmatrix}, \\
    \ket{\phi_3} &= \alpha_3 \begin{pmatrix} \frac{\sqrt{2}(U-V-X)}{4J} \\ 1 \\ 1 \end{pmatrix}.
\end{aligned}
\end{equation}
The corresponding eigenvalues are:
\begin{equation}
\begin{aligned}
    \lambda_1 &= U, \\
    \lambda_2 &= \frac{1}{2} \left(U + V - X\right), \\
    \lambda_3 &= \frac{1}{2} \left(U + V + X\right),
\end{aligned}
\end{equation}
where $X = \sqrt{16J^2 + (U - V)^2}$, and $\alpha_2, \alpha_3$ are normalization constants.

For an initial state $\ket{\psi(0)}=\ket{1;1}_L$, the evolution is given by
\begin{align}
    \ket{\psi(t)} = \frac{\sqrt{2}J}{X} \left(e^{-i\lambda_2 t} \ket{\phi_2} - e^{-i\lambda_3 t} \ket{\phi_3} \right),
\end{align}
where $\ket{\phi_2}$ and $\ket{\phi_3}$ are eigenstates of the Hamiltonian $\hat{H}_4$. To ensure that the state $\ket{\psi(t)}$ remains in the computational subspace $\mathcal{H}_{\rm C}$ by the end, we impose the additional condition:
\begin{equation}\label{eq: condition2}
    e^{-i\lambda_2 T_{\rm CP}} = e^{-i\lambda_3 T_{\rm CP}},
\end{equation}
which implies the constraint:
\begin{equation}
\lambda_2 T_{\rm CP} = \lambda_3 T_{\rm CP} - 2m\pi, \quad m \in \mathbb{N}^*.
\end{equation}

Using the expression for the gate time $T_{\rm CP}$ (given in Eq.~\eqref{eq: CP time}), we obtain the following restriction on the parameters:
\begin{equation}\label{eq: restriction of CP}
    \frac{U - V_{\rm CP}}{J_{\rm CP}} = \pm \sqrt{m^2 - 16}, \quad m \geqslant 4.
\end{equation}
Here, we have used the simplified notation \( X = m|J| \). From this, we express the normalization coefficients as:
\begin{align*}
    \alpha_2^2 &= \frac{4}{m^2 - m\sqrt{m^2 - 16}}, \\
    \alpha_3^2 &= \frac{4}{m^2 + m\sqrt{m^2 - 16}}.
\end{align*}
The accumulated phase at the gate time $T_{\rm CP}$ is then given by:
\begin{equation}\label{eq: phase of CP}
    \varphi_m = -\pi \left(\frac{2V_{\rm CP}}{J_{\rm CP}} - \sqrt{m^2 - 16} - m\right).
\end{equation}

Thus, the unitary evolution operator $\hat{U}_{\rm CP}(T_{\rm CP})$ can be expressed as a block diagonal matrix, with contributions from the computational subspace and the orthogonal subspace $\mathcal{H}_{\perp}$:
\begin{equation}
    \hat{U}_{\rm CP}(T_{\rm CP}) = \hat{U}_{\rm CP} \oplus \hat{U}_{\perp} = \mathrm{diag}\left(1, 1, 1, \exp(i\varphi_m)\right) \oplus \hat{U}_{\perp},
\end{equation}
where $\hat{U}_{\perp}$ acts on $\mathcal{H}_{\perp}$, the non-computational subspace.
For integers \( m \geqslant 4 \), the desired phase \( \varphi_m \) can be achieved by selecting an appropriate value for \( m \) and tuning the ratio \( V_{\rm CP} / J_{\rm CP} \).

Next, we apply this framework to a practical parameter set for the CZ gate. Typically, the effective ZZ strength \( V_{\rm CP} \) is much smaller than the coupling strength \( J_{\rm CP} \). We choose \( m = 8 \), leading to a ratio \( V_{\rm CP} / J_{\rm CP} \approx -0.036 \). Under this condition, the restriction in Eq.~\eqref{eq: restriction of CP} simplifies to:
\begin{equation}
U = V_{\rm CP} \pm 4\sqrt{3}J_{\rm CP}.
\end{equation}
Since the anharmonicity of a transmon is negative, we select the minus sign, yielding \( U \approx -6.964J_{\rm CP} \). We adopt \( J_{\rm CP} \) as the unit of energy, and for a typical coupling strength \( J_{\rm CP}/2\pi = 40 \, \text{MHz} \), we estimate:
\begin{equation}
V_{\rm CP}/2\pi \approx -1.44 \, \text{MHz}, \quad U/2\pi \approx -278.6 \, \text{MHz}.
\end{equation}
These values lie within the practical regime for the transmon and tunable coupling scheme, corresponding to a gate time of \( T_{\rm CP} = 25 \, \text{ns} \).

The small ratio \( V_{\rm CP} / J_{\rm CP} \) can be increased by employing an alternative coupling scheme, as proposed in Ref.~\cite{kounalakis2018tuneable}, where the coupler consists of a parallel combination of a capacitor and a Josephson junction, potentially allowing for \( V/J \) ratios greater than one.

\begin{figure}[t]
	\centering
    \subfigure{\begin{minipage}[b]{0.45\columnwidth}
        \centering
            \begin{overpic}[scale=0.3]{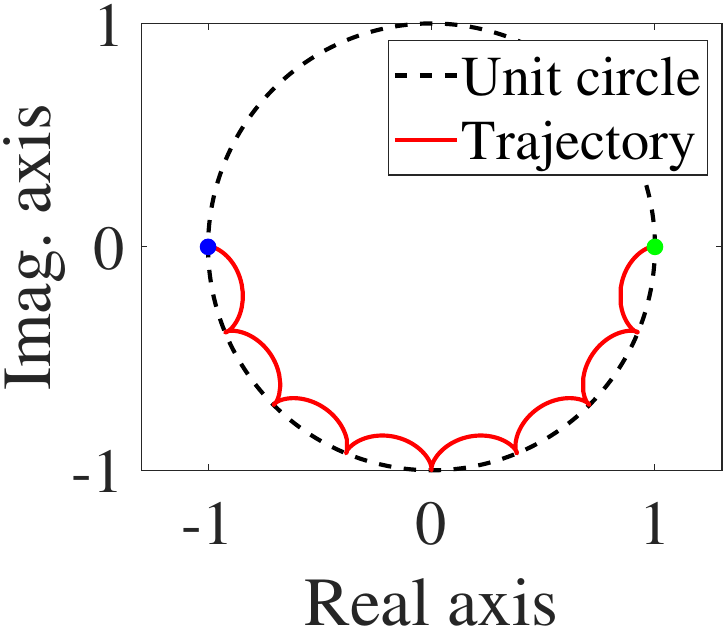}
            \put(0,95){\textbf{(a)}}
        \end{overpic}
    \end{minipage}
    \label{fig: phase 1}
}
\hspace{0.0\columnwidth} 
 \subfigure{\begin{minipage}[b]{0.45\columnwidth}
        \centering
            \begin{overpic}[scale=0.3]{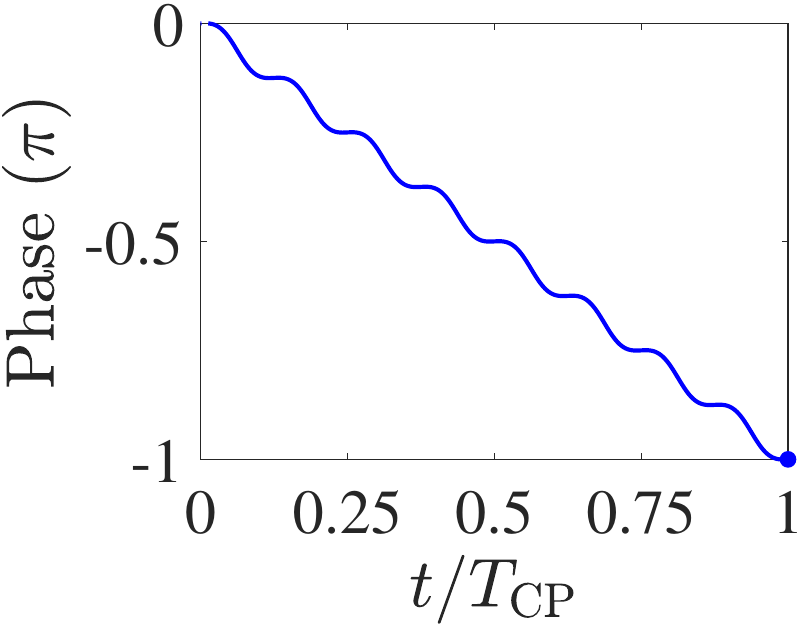}
            \put(0,85){\textbf{(b)}}
        \end{overpic}
    \end{minipage}
    \label{fig: phase 2}
}	
	\caption{
    \textbf{Evolution of the state $\ket{1;1}_L$ for implementing a CZ gate.}
    (a) Coefficient trajectory (solid red line) of $\ket{1;1}_L$ in the complex plane. The black dashed curve shows the unit circle. The $x$-axis and $y$-axis represent the real and imaginary parts, respectively. The state starts at $(1, 0)$ (green dot) and ends at the target point $(-1, 0)$ (blue dot).
    (b) Phase evolution. The phase decreases continuously with small oscillations, reaching $-\pi$ at the gate time.
    }
\label{fig: phase}
\end{figure}

We show the population evolution for the CZ gate in Fig.~\ref{fig: CZgate}(b), using parameters \( m = 8 \), \( V_{\rm CP} = (-7/2 + 2\sqrt{3}) J_{\rm CP} \), and \( U = -(7/2 + 2\sqrt{3}) J_{\rm CP} \). The initial states are chosen from the computational subspace \( \mathcal{H}_{\rm C} \), with \( \ket{\psi(0)} = \ket{0;1}_L \) in Fig.~\ref{fig: CZgate}(b1) or \( \ket{1;1}_L \) in Fig.~\ref{fig: CZgate}(b2). The population of each computational basis state returns to unity at the gate time \( T_{\rm CP} \), and the population of \( \ket{1;1}_L \) remains large throughout the evolution, maximizing the effect of the ZZ interaction.

We further present the numerical evolution of the basis state $\ket{1;1}_L$ in Fig.~\ref{fig: phase}. It evolves under the Hamiltonian $H_4$. 
In Fig.~\ref{fig: phase 1}, we show the trajectory of the complex coefficient of $\ket{1;1}_L$. Initialized at $(x=1,y=0)$ (green dot), the state evolves into the target point $(-1,0)$ (blue dot) at the gate time $T_{\rm CP}$, corresponding to $\ket{\psi(T_{\rm CP})} = -\ket{1;1}_L$. 
Fig.~\ref{fig: phase 2} illustrates the phase accumulation of the coefficient of $\ket{1;1}_L$, which decreases with small oscillations, reaching $-\pi$ at the gate time.


In conclusion, both analytical calculations and numerical simulations verify that the computational basis states return to themselves, while the state \( \ket{1;1}_L \) acquires an additional phase of \( -\pi \), confirming the successful implementation of the CZ gate.

\subsection{CPhase gate for longitudinal connections}
\begin{figure}[t]
	\centering
	\includegraphics[width=0.9\columnwidth]{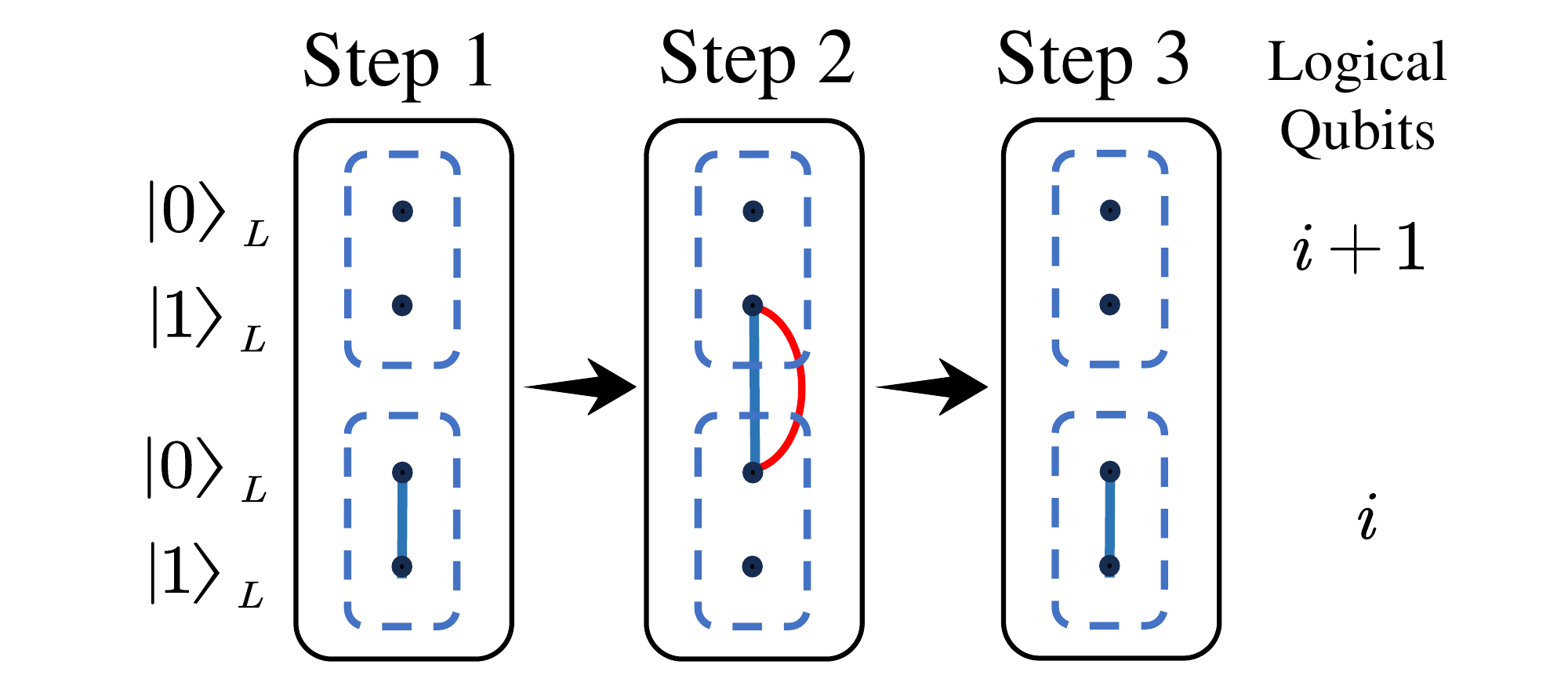}
	\caption{
\textbf{Implementation of the CPhase gate with longitudinal connected logical qubits.} 
Since the coupling between logical qubits is restricted to $\ket{0_i}_L$ and $\ket{1_{i+1}}_L$, a basis transformation is required to swap $\ket{0_i}_L$ into $\ket{1_i}_L$. We first perform an X gate on the $i$-th logical qubit. The second step follows the same procedure as outlined in the previous subsection. Finally, another X gate is applied to revert the basis back to its original form.
}
	\label{fig: CZlong}
\end{figure}

As shown in Fig.~\ref{Fig: Illustrating}(a), longitudinally connected logical qubits are restricted to a single coupling between sites \( (i,0) \) and \( (i+1,1) \), limiting the available two-qubit gate operations. However, similar to the transverse connection, the CPhase gate can still be implemented by utilizing a single coupling, with additional single-qubit operations. This process is illustrated in Fig.~\ref{fig: CZlong}.

The implementation proceeds in three steps. First, a local \( X \)-gate is applied to the $i$-th logical qubit, transforming \( \ket{1_i}_L \) into \( \ket{0_i}_L \). This operation takes a time \( t_1 = \pi / (2J_X) \), where \( J_X \) is the coupling strength for the \( X \)-gate. The second step involves applying a CPhase operation, following the procedure from the previous subsection, with a duration \( t_2 = 2\pi / |J_{\text{CP}}| \), where \( J_{\text{CP}} \) is the coupling strength for the CPhase gate. After this step, the state \( \ket{0_i1_{i+1}}_L \) acquires an additional phase. Finally, a second \( X \)-gate is applied to return the basis state to its original form, taking a time \( t_3 = t_1 \).

Thus, the total gate time for the longitudinal CPhase gate is:
\begin{equation}
    T_{\text{CPL}} = \pi \left( \frac{1}{J_X} + \frac{2}{J_{\text{CP}}} \right).
\end{equation}
It is important to note that the coupling strengths vary for different operations. The \( X \)-gate demands a zero ZZ interaction strength $V_X=0$, while the CPhase gate requires a specific ratio between \( V_{\rm CP} \) and \( J_{\rm CP} \) to ensure the correct phase accumulation.

\subsection{iSWAP gate for transverse connections}
\begin{figure}[t]
	\centering
	\includegraphics[width=\columnwidth]{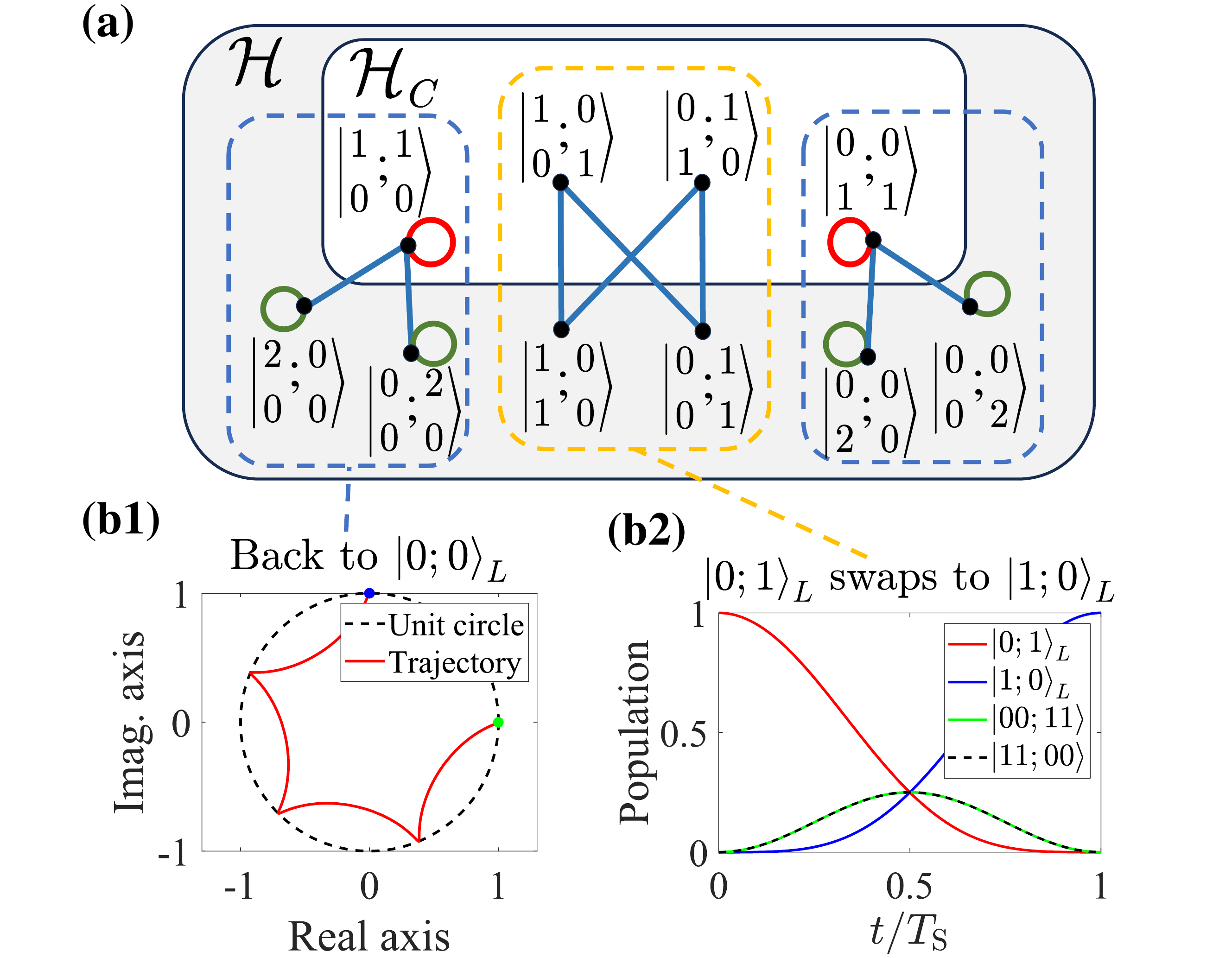}
	\caption{
    \textbf{Dual graph and evolution of the iSWAP Gate on dual-rail encoded qubits.}
(a) The graph illustrates the couplings and interactions between Fock states under the Hamiltonian \( \hat{H}_{\text{S}} \). It consists of 10 vertices, each representing a basis state in the Fock space \( \mathcal{H} \) of two bosons distributed over four sites. The four selected states span the computational subspace \( \mathcal{H}_{\rm C} \). Blue edges correspond to allowed couplings, red self-loops represent ZZ interactions, and green self-loops correspond to on-site interactions.
(b1) Coefficient trajectory of the state \( \ket{0;0}_L \) on the complex plane. The state begins at \( (1,0) \) (green dot) and evolves to \( (0,1) \) (blue dot) at the gate time \( T_{\rm S} \). At the final time, the population of \( \ket{0;0}_L \) is unity, and the accumulated phase is \( i \).
(b2) Evolution of the state \( \ket{0;1}_L \) swapping to \( \ket{1;0}_L \). The $x$-axis shows the evolution time in units of \( T_{\rm S} \), while the $y$-axis indicates the population of the involved states. Despite the non-computational states being populated during the evolution, their populations diminish by the time \( T_{\rm S} \), confirming a perfect swap process.
}
	\label{fig: iSWAPgate}
\end{figure}

The iSWAP gate is another maximally entangling gate and, when combined with single-qubit gates, forms a universal gate set for quantum computation~\cite{divincenzo1995two}. In the transverse connection mode, the iSWAP gate can readily be implemented using the ZZ interaction. To achieve this, we consider implementing an iSWAP operation with a global phase \(i\), denoted by \( \hat{U}_{\rm S} \), whose matrix representation is:
\begin{equation}
\hat{U}_{\rm S} = i \cdot \hat{U}_{\text{iSWAP}} = 
i \begin{pmatrix}
1 & 0 & 0 & 0 \\
0 & 0 & i & 0 \\
0 & i & 0 & 0 \\
0 & 0 & 0 & 1
\end{pmatrix}
= 
\begin{pmatrix}
i & 0 & 0 & 0 \\
0 & 0 & -1 & 0 \\
0 & -1 & 0 & 0 \\
0 & 0 & 0 & i
\end{pmatrix}.
\end{equation}

A straightforward implementation of this gate involves simultaneously opening the couplings between adjacent logical qubits \( \ket{x_i}_L \) and \( \ket{x_{i+1}}_L \) for each \( x \in \{0,1\} \), as shown in Fig.~\ref{Fig: Illustrating}(b). The additional global phase \(i\) is compensated by applying a frequency shift \( \mu_{\rm S} \hat{n} \) to each of the four physical qubits until their phase accumulations reach \( -\pi/2 \). 

For simplicity, we focus on the \( \hat{U}_{\rm S} \) operation. Denoting the coupling and interaction strengths as \( J_{\rm S} \) and \( V_{\rm S} \), the system Hamiltonian is:
\begin{equation}
    \begin{aligned}
        \hat{H}_{\text{S}} =& -J_{\rm S} \left( \hat{c}^\dagger_{i,0} \hat{c}_{i+1,0} + \hat{c}^\dagger_{i,1} \hat{c}_{i+1,1} + \text{H.c.} \right)\\
        &+ V_{\rm CP} \left( \hat{n}_{i,0} \hat{n}_{i+1,0} + \hat{n}_{i,1} \hat{n}_{i+1,1} \right) + \hat{H}_0,
    \end{aligned}
\end{equation}
where \( \hat{H}_0 \) is the default Hamiltonian. The evolution operator is \( \hat{U}_{\rm S}(t) = \exp\left( -i \hat{H}_{\rm S} t \right) \). As in previous subsections, we derive conditions for unitary evolution within the computational subspace \( \mathcal{H}_{\rm C} \) and examine the system's action on the computational basis states, as shown in Fig.~\ref{fig: iSWAPgate}(a). The dashed boxes can be classified into two scenarios, marked by orange or blue colors.

 
In the first scenario, the Hamiltonian \( \hat{H}_{\rm S} \) couples the computational states \( \ket{10;01} \) and \( \ket{01;10} \), along with non-computational states \( \ket{11;00} \) and \( \ket{00;11} \). Their dynamics are governed by two simultaneous Rabi oscillations. Assuming an initial state \( \ket{10;01} \), the evolutions for the upper and lower rows, are respectively:
\begin{align*}
    \hat{U}_{\rm S}(t) \ket{10} &= \cos(J_{\rm S} t) \ket{10} + i \sin(J_{\rm S} t) \ket{01},\\
    \hat{U}_{\rm S}(t) \ket{01} &= \cos(J_{\rm S} t) \ket{01} + i \sin(J_{\rm S} t) \ket{10}.
\end{align*}
Thus, the evolution of the two-qubit state \( \ket{10;01} \) becomes:
\begin{align*}
    \hat{U}_{\rm S}(t) \ket{10;01} =& \cos^2(J_{\rm S} t) \ket{10;01} - \sin^2(J_{\rm S} t) \ket{01;10}\\
    &+ \frac{i}{2} \sin(2J_{\rm S} t) \left( \ket{11;00} + \ket{00;11} \right).
\end{align*}

To achieve a perfect swap of the computational states \( \ket{10;01} \) and \( \ket{01;10} \), the gate time must satisfy:
\begin{equation}
T_{\rm S} = \frac{(2k+1) \pi}{2 J_{\rm S}}, \quad k \in \mathbb{N}.
\end{equation}
At this time, the evolution is:
\begin{equation}
\hat{U}_{\rm S}(T_{\rm S}) \ket{10;01} = - \ket{01;10}, \ \  \hat{U}_{\rm S}(T_{\rm S}) \ket{01;10} = - \ket{10;01}.
\end{equation}
Due to the symmetry of \( \hat{H}_{\rm S} \), the evolutions and restrictions for the state $\ket{01;10}$ are identical.

In the second scenario, the evolution of the state $\ket{10;10}$ is the same as $\ket{01;01}$. 
We focus on \( \ket{10;10} \), which is coupled to the non-computational states \( \ket{20;00} \) and \( \ket{00;20} \). In this subspace, the Hamiltonian matrix is the same as in Eq.~\eqref{eq: 3_d_Hamiltonian}. 
For \( \hat{U}_{\rm S} \) to return \( \ket{10;10} \) to itself at \( T_{\rm S} \) while accumulating an additional phase \(i\), we impose two conditions on the system parameters.


The first restriction ensures that the state returns to itself at \( T_{\rm S} \). This condition is:
\begin{equation}
\frac{U - V_{\rm S}}{J_{\rm S}} = -4 \sqrt{\frac{m^2}{(2k+1)^2} - 1}, \quad 2k+1 < m \in \mathbb{Z}.
\end{equation}


The second restriction ensures that the accumulated phase is equal to $i$. This condition is:
\begin{equation}
\frac{V_{\rm S}}{J_{\rm S}} = 2 \sqrt{\frac{m^2}{(2k+1)^2} - 1} + \frac{2m - 4n - 1}{2k+1}, \quad n \in \mathbb{Z}.
\end{equation}


A possible set of parameters is:
\begin{equation}
m = 4, \quad n = 3, \quad k = 2.
\end{equation}
This choice leads to:
\begin{equation}
V_{\rm S} \approx 0.1 J_{\rm S}, \quad U \approx -3.4 J_{\rm S},
\end{equation}
and the total iSWAP gate time is:
\begin{equation}
T_{\rm S} = \frac{5\pi}{2 J_{\rm S}}.
\end{equation}
Within this gate time, the detuning on each physical qubit should be \( \mu_{\rm S} = 0.2 J_{\rm S} \).


In Fig.~\ref{fig: iSWAPgate}(b), we show the numerical evolution of the basis states during the iSWAP gate operation with the parameters described above.
In Fig.~\ref{fig: iSWAPgate}(b1), we present the trajectory of the complex coefficient for \( \ket{0;0}_L \) (equivalently \( \ket{1;1}_L \)). Starting with \( \ket{\psi(0)} = c(0) \ket{0;0}_L \) and \( c(0) = 1 \) (green dot), the evolution leads back to \( \ket{0;0}_L \) at \( T_{\rm S} \), with \( c(T_{\rm S}) = i \) (blue dot), completing the desired operation.
In Fig.~\ref{fig: iSWAPgate}(b2), we show the swapping dynamics between \( \ket{0;1}_L \) and \( \ket{1;0}_L \). The $x$-axis represents the evolution time, and the $y$-axis indicates the population of the involved states. The populations follow \( \cos^4(J_{\rm S} t) \) and \( \sin^4(J_{\rm S} t) \) curves during the evolution, demonstrating smooth transitions between the two states. This suggests potential robustness against fluctuations in the coupling strength \( J_{\rm S} \), which could be advantageous for practical implementations.

\subsection{CCPhase Gate for Transverse Connections}

\begin{figure}[t]
	\centering
	\includegraphics[width=\columnwidth]{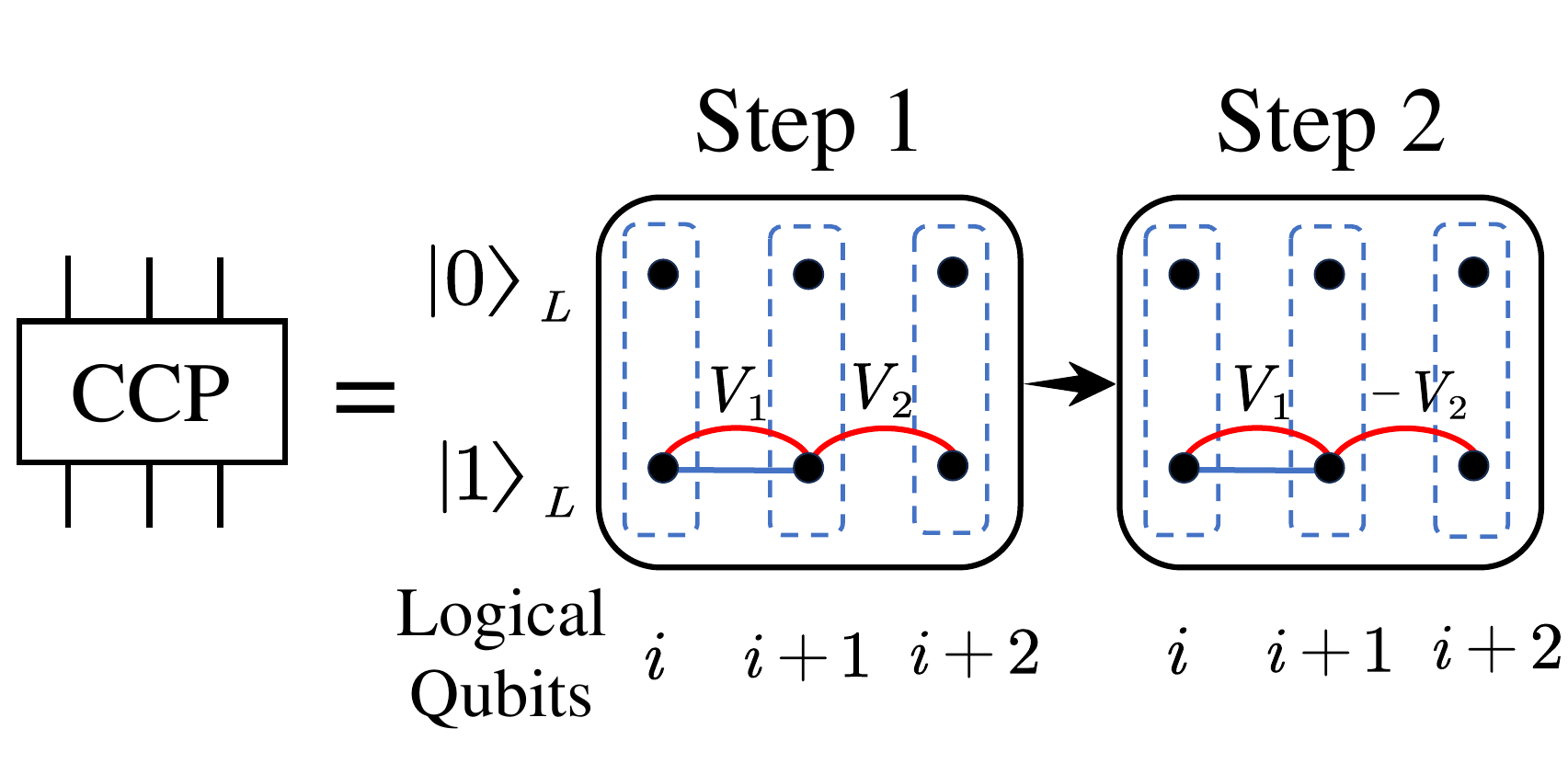}
	\caption{
\textbf{Implementation of the CCPhase Gate with three neighboring logical qubits.} 
The gate operation consists of two steps. In each step, a CZ gate is applied to the first two qubits in the \( x=1 \) row while alternately enabling ZZ interactions \( V_2 \) and \( -V_2 \) between the latter two qubits. This ensures zero phase accumulation for \( \ket{1;1;0}_L \), \( \ket{1;0;1}_L \), and \( \ket{0;1;1}_L \), while a nonzero phase is accumulated exclusively on \( \ket{1;1;1}_L \).}
	\label{fig: CCP}
\end{figure}

The CCPhase gate is a crucial operation for three neighboring logical qubits, requiring precise control of phase accumulation across computational basis states. This subsection introduces the implementation of the CCPhase gate using a two-step approach, ensuring phase cancellation for all states except \( \ket{1;1;1}_L \), which accumulates the desired phase shift. We illustrate the quantum walk dynamic diagram for implementing the CCPhase gate in Fig.~\ref{fig: CCP}. In each step, a CZ gate is applied to the first two qubits in the \( x=1 \) row while alternately enabling ZZ interactions \( V_2 \) and \( -V_2 \) between the latter two qubits. 

The dynamics are confined to states involving at least one walker in the \( x=1 \) row. 
For states with one walker in the \( x=1 \) row (\( \ket{1;0;0}_L \), \( \ket{0;1;0}_L \), and \( \ket{0;0;1}_L \)),
the Rabi oscillation between the first two states requires a time identical to $T_{\text{CP}}$, while the last state is stationary. Thus, the two-step approach sets up the gate time
\( T_{\text{CCP}}=2T_{\text{CP}} \).

There are three two-walker states, \( \ket{1;0;1}_L \), \( \ket{0;1;1}_L \), and \( \ket{1;1;0}_L \). The first two states are coupled due to the coupling between $\ket{1_i}_L$ and $\ket{1_{i+1}}_L$, while the dynamics of the third state are the same as in CPhase gate. 

The Hamiltonian acting on the states \( \ket{1;0;1}_L \) and \( \ket{0;1;1}_L \) is described as
\begin{equation}
\hat{H}_1 = 
\begin{pmatrix}
0 & -J \\
-J & V_2 \\
\end{pmatrix}.
\end{equation}
The eigenvalues and eigenstates are:
\begin{equation}
\begin{aligned}
\lambda_{1,2} &= \frac{1}{2} \left(V_2 \pm \sqrt{4 J^2 + V_2^2}\right), \\
\ket{\phi_{1,2}} &= \alpha_{1,2} 
\begin{pmatrix}
\frac{V_2 \pm \sqrt{4 J^2 + V_2^2}}{2 J}, & 1
\end{pmatrix}^\mathrm{T}.
\end{aligned}
\end{equation}
To ensure \( \ket{1;0;1}_L \) and \( \ket{0;1;1}_L \) return to their original states after \( T_{\text{CP}} \), the interaction strength must satisfy:
\begin{equation}
\frac{V_2}{J} = \pm \sqrt{k^2 - 4}, \quad k \geqslant 2, \quad k \in \mathbb{N}.
\end{equation}
Under this restriction, the accumulated phase on \( \ket{1;0;1}_L \) reads
\begin{equation}
\varphi_{101}(V_2) = -\pi \left(\frac{V_2}{J} + k\right).
\end{equation}
Thus, the phase accumulation cancels over two steps:
\begin{equation}
\varphi_{101}(V_2) + \varphi_{101}(-V_2) = 0.
\end{equation}
These arguments apply to the state \( \ket{0;1;1}_L \) as well.

On the other hand, the dynamics of \( \ket{1;1;0}_L \) mimic those of \( \ket{1;1}_L \) in the CPhase gate. To prevent leakage, the condition is:
\begin{equation}
\frac{U - V_1}{J} = \pm \sqrt{m^2 - 16}, \quad m \geqslant 4, \quad m \in \mathbb{Z}.
\end{equation}
Zero accumulated phase requires:
\begin{equation}
\frac{V_1}{J} = \frac{1}{2} \left(\sqrt{m^2 - 16} + m + k\right), \quad k \in \mathbb{Z}.
\end{equation}

Finally, for three walkers, \( \ket{1;1;1}_L \) couples to non-computational states \( \ket{2;0;1}_L \) and \( \ket{0;2;1}_L \), described by the Hamiltonian:
\begin{equation}
\hat{H}_2 =
\begin{pmatrix}
V_1 + V_2 & -\sqrt{2}J & -\sqrt{2}J \\
-\sqrt{2}J & U & 0 \\
-\sqrt{2}J & 0 & U + 2V_2
\end{pmatrix}.
\end{equation}
For small \( V_2 / J \), we find the resulting phase accumulation is negligibly small by applying perturbative calculations. When \( V_2 / J \) is large, we can search possible parameters to ensure \( \ket{1;1;1}_L \) return to itself after $T_{\text{CP}}$, while $V_2$ being around the value 
\begin{equation}
\frac{V_2}{J} = \sqrt{k^2 - 4}, \quad k \geqslant 3.
\end{equation}

The parameters achieving \( F \geqslant 0.99 \) fidelity are:
\begin{equation}
\begin{aligned}
    &\frac{V_2}{J} \approx 2\sqrt{3}, \quad m \geqslant 14, \quad \frac{V_1}{J} \approx 0.2, \\
    &U = V_1 - \sqrt{m^2 - 16}.
\end{aligned}
\end{equation}
The large values of \( V_1 \) and \( V_2 \) require a coupling scheme as proposed in Ref.~\cite{kounalakis2018tuneable}. We find that the maximum phase on \( \ket{1;1;1}_L \) is \( -0.086\pi \), with a total gate time of \( T_{\text{CCP}} = 2T_{\text{CP}} \).

\section{Noise Analysis on the CZ Gate}

In this section, we analyze the effects of various noise sources on the performance of the CZ gate in superconducting circuits. The CZ gate serves as a representative example due to its practical implementation within a single step under realistic parameters. The primary noise sources considered are:

1. \textbf{Dephasing:} Fluctuations in qubit energy levels lead to loss of coherence and degrade gate fidelity.

2. \textbf{Relaxation:} Energy relaxation from the excited to the ground state during gate operation reduces the population of the computational subspace, introducing errors.

3. \textbf{Inaccurate Hamiltonian Parameters:} Deviations in coupling strength \( J \) and interaction strength \( V \) can cause the system's evolution to diverge from the ideal unitary operation.

4. \textbf{Detuning Errors:} Misalignment in detuning between adjacent logical qubit sites affects phase accumulation, potentially inducing leakage into non-computational subspaces.

Strategies to mitigate these effects include parameter optimization, error correction, and noise-resilient
designs for superconducting circuits. We will quantify the impact of these noise sources on the CZ gate. The Hamiltonian parameters for the CZ gate implementation are set as follows: \( m = 8 \), \( V_{\rm CP} = (-7/2 + 2\sqrt{3}) J_{\rm CP} \), \( U = -(7/2 + 2\sqrt{3}) J_{\rm CP} \), and \( \mu = 0 \).

\subsection{Dephasing and Relaxation}

The effects of {dephasing} and {relaxation} are analyzed using the Lindblad master equation, which governs the evolution of the system's density matrix \( \rho \):
\begin{align}
    \frac{d\rho}{dt} = -\frac{i}{\hbar} [\hat{H}, \rho] + \sum_k \left( \hat{L}_k \rho \hat{L}_k^\dagger - \frac{1}{2} \left\{ \hat{L}_k^\dagger \hat{L}_k, \rho \right\} \right),
\end{align}
where \( \hat{H} \) is the system Hamiltonian, and \( \hat{L}_k \) are Lindblad operators describing interactions between the quantum system and its environment.

Dephasing noise leads to coherence loss without altering population distributions in the computational basis. For a homogeneous dephasing rate \( \gamma \), the Lindblad operator associated with each computational basis state \( \ket{k} \) is:
\begin{equation}
    \hat{L}_k = \sqrt{\gamma} \ket{k}\bra{k}.
\end{equation}
Relaxation involves energy dissipation and is characterized by the relaxation rate \( \Gamma \). The Lindblad operators for relaxation on each physical qubit are given by:
\begin{equation}
    \hat{L}_{\downarrow,i,x} = \sqrt{\Gamma} \sum_{k=0,1} \ket{k} \bra{k+1},
\end{equation}
where \( i \) and \( x \) index the qubit in a \( 2 \times n \) array of physical qubits.

We assume a relaxation time \( T_1 = 100 \, \upmu\text{s} \), corresponding to \( \Gamma = J / 5000 \) when the coupling strength \( J / 2\pi = 50 \, \text{MHz} \). The dephasing rate is set to \( \gamma = \Gamma \), consistent with typical experimental conditions. Notably, dual-rail qubits encoded in resonantly coupled transmons have demonstrated coherence times exceeding milliseconds~\cite{levine2024demonstrating}, highlighting their potential for quantum information processing.

\begin{figure}[t]     
	\centering
\subfigure{\begin{minipage}[b]{\columnwidth}
        \centering
            \begin{overpic}[scale=0.4]{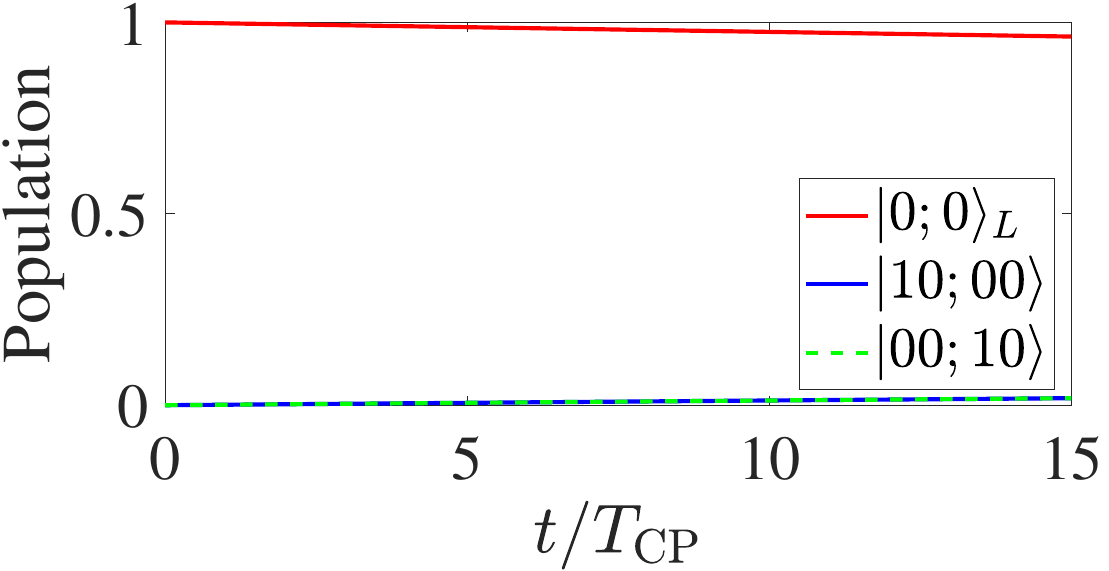}
            \put(0,55){\textbf{(a)}}
        \end{overpic}
    \end{minipage}
    \label{noise 1}
    }
\subfigure{\begin{minipage}[b]{\columnwidth}
        \centering
            \begin{overpic}[scale=0.4]{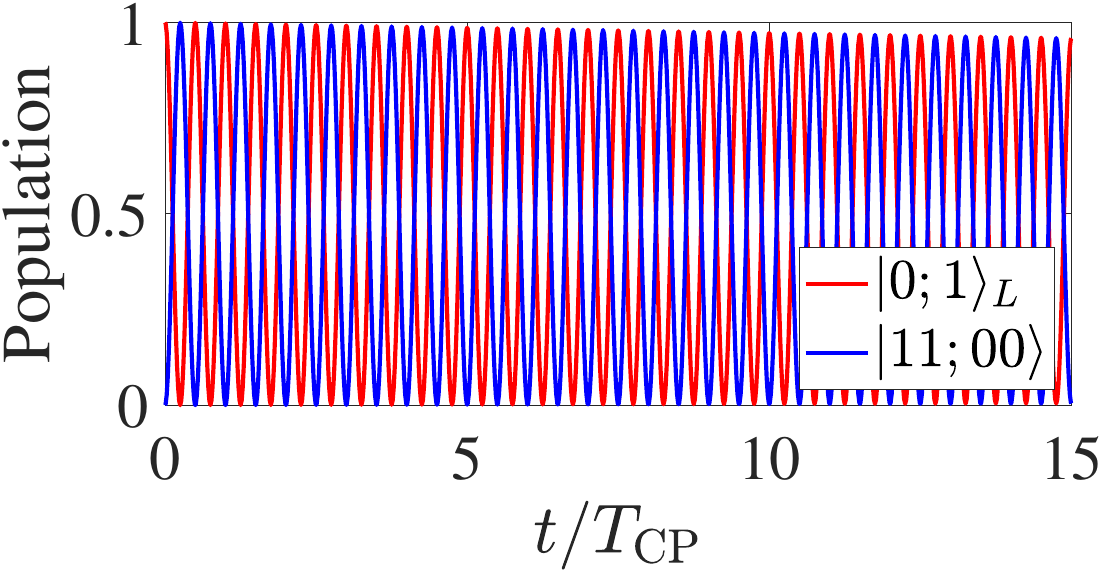}
            \put(0,55){\textbf{(b)}}
        \end{overpic}
    \end{minipage}
    \label{noise 2}
}
\subfigure{\begin{minipage}[b]{\columnwidth}
        \centering
            \begin{overpic}[scale=0.4]{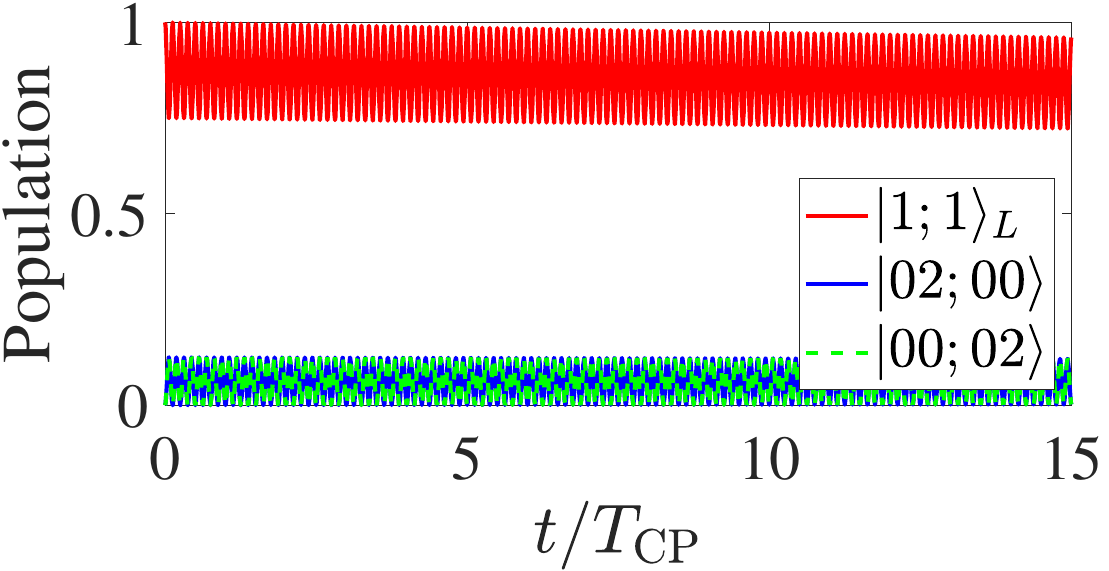}
            \put(0,55){\textbf{(c)}}
        \end{overpic}
    \end{minipage}
    \label{noise 3}
    }    
    
    \caption{\textbf{Population dynamics of basis states under relaxation and dephasing Noise.} The computational basis states are represented by the red lines. (a) The population of the state $\ket{0;0}_L$ decreases due to relaxation, while the populations of the leakage states $\ket{10;00}$ (blue) and $\ket{00;10}$ (green dashed) increase. (b) The population of the computational state $\ket{0;1}_L$ decreases, accompanied by a decay in the leakage state $\ket{11;00}$ (blue). (c) The population of the computational state $\ket{1;1}_L$ decreases due to relaxation, whereas the leakage states $\ket{02;00}$ (blue) and $\ket{00;02}$ (green dashed) show negligible decay.}
    \label{fig: dephasing and relaxation noise}
\end{figure}

The numerical simulation results are presented in Fig.~\ref{fig: dephasing and relaxation noise}, showcasing the effects of dephasing and relaxation noise on the population dynamics of selected initial states: $\ket{0;0}_L$, $\ket{0;1}_L$, and $\ket{1;1}_L$. The simulations were performed using the \texttt{ode45} solver in MATLAB to solve the Lindblad master equation, with the evolution time extending up to 15 times the CPhase gate duration.

In Fig.~\ref{noise 1}, the initial state $\ket{0;0}_L$ experiences a population decay to 0.963 by the end of the evolution, accompanied by a population growth in the leakage states $\ket{10;00}$ and $\ket{00;10}$, which together account for 0.037 of the total population. These leakage events, which can be detected and converted into erasure errors, highlight the importance of error mitigation strategies.

Figure~\ref{noise 2} shows the evolution of the state $\ket{0;1}_L$, which exhibits Rabi oscillations with the leakage state $\ket{11;00}$. By the end of the evolution, the population of $\ket{0;1}_L$ decreases to 0.959, while the leakage states $\ket{10;00}$ and $\ket{00;01}$ acquire populations of 0.0183 and 0.0137, respectively. This behavior underscores the need for an effective error correction scheme tailored to this subspace.

Finally, Fig.~\ref{noise 3} examines the dynamics of the initial state $\ket{1;1}_L$, which decays to a population of 0.960. The associated leakage states $\ket{02;00}$ and $\ket{00;02}$, critical for phase accumulation, exhibit negligible population changes, with their combined population decreasing marginally from 0.125 to 0.121. The remaining population loss is primarily accounted for by the leakage states $\ket{00;01}$ and $\ket{01;00}$, which together contribute 0.035. These results suggest that the phase deviation introduced by the leakage states $\ket{02;00}$ and $\ket{00;02}$ is minor, maintaining the robustness of the phase-sensitive operations.

\subsection{Impact of Parameter Deviations on CZ Gate}
\begin{figure}[t]
	\centering
\subfigure{\begin{minipage}[t]{\columnwidth}
        \centering
            \begin{overpic}[scale=0.13]{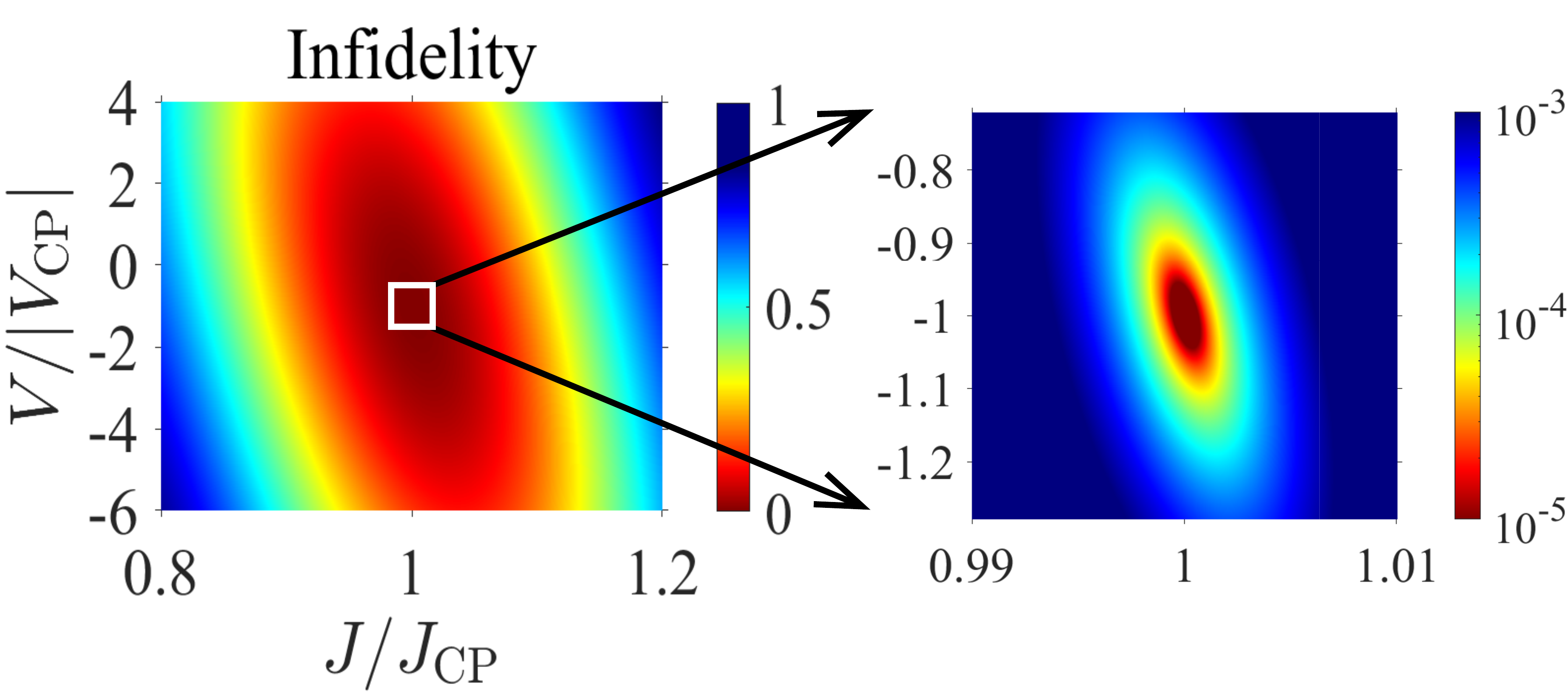}
            \put(0,45){\textbf{(a)}}
        \end{overpic}
    \end{minipage}
    \label{fig: infidelity deviation}
    }
\subfigure{\begin{minipage}[t]{\columnwidth}
        \centering
            \begin{overpic}[scale=0.13]{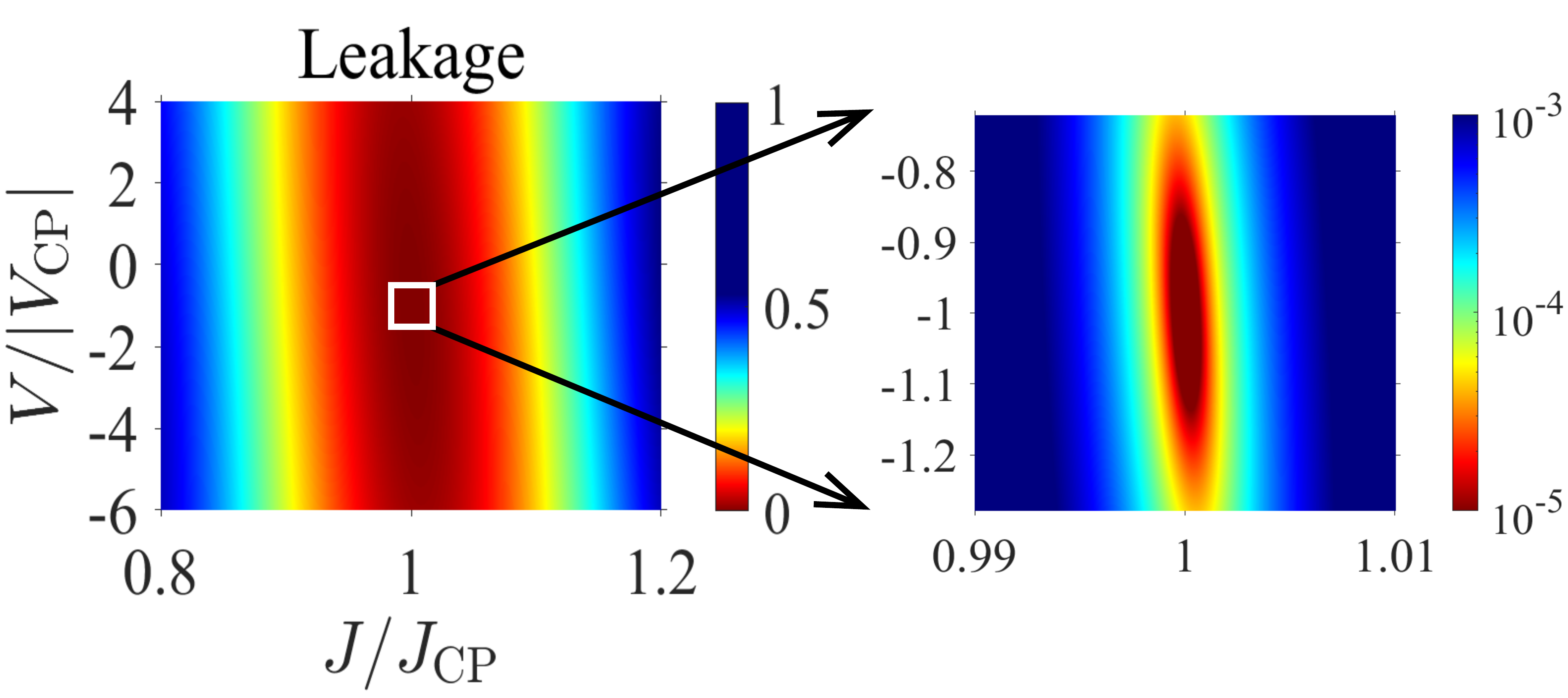}
            \put(0,45){\textbf{(b)}}
        \end{overpic}
    \end{minipage}
    \label{fig: leakage deviation}
}   
    \caption{\textbf{Impact of parameter imperfections on CZ gate performance, shown via (a) fidelity and (b) leakage.} 
    The $x$-axis represents deviations in the coupling strength, measured in units of $J_{\text{CP}}$, while the $y$-axis corresponds to the ZZ interaction strength, expressed in units of $|V_{\text{CP}}|$. 
    The right panels provide a magnified view of the region near the optimal parameters $J = J_{\text{CP}}$ and $V = V_{\text{CP}}$. Owing to the large value of $U$, variations in $J$ have a more pronounced effect compared to deviations in $V$. Nevertheless, the impact of both parameters remains within acceptable bounds under typical experimental noise conditions.}
    \label{fig: infidelity and leakage}
\end{figure}

Understanding the robustness of the CZ gate to parameter deviations is crucial for reliable quantum computation. Variations in the coupling strength \( J \) and interaction strength \( V \), caused by flux noise or crosstalk, can alter the effective Hamiltonian and introduce leakage or fidelity loss. This section examines the impact of such deviations and quantifies their effects on gate performance, including leakage and fidelity.

The coupler frequency determines both \( J \) and \( V \), making the system sensitive to flux noise. Additionally, crosstalk can induce residual ZZ interactions~\cite{mundada2019suppression}. Typical deviations in superconducting circuits include shifts in the coupling strength of approximately \( 5 \, \text{kHz} \times 2\pi \) and residual ZZ interactions up to \( 10 \, \text{kHz} \times 2\pi \). These deviations can prevent the system from fully returning to the computational subspace at the end of the gate operation, leading to true leakage events.

The fidelity of the CZ gate is defined using the expression~\cite{pedersen2007fidelity}:
\begin{equation}
    F = \frac{1}{n(n+1)} \left[ \Tr(MM^{\dagger}) + \left| \Tr(M) \right|^2 \right],
\end{equation}
where \( n \) is the dimension of the computational subspace, \( M = PU_{\rm CZ}^{\dagger}UP \), and \( U_{\rm CZ} = \mathrm{diag}(1, 1, 1, -1) \) is the ideal CZ gate unitary. Here, \( U \) is the unitary operation under noise, and \( P \) projects onto the computational subspace \( \mathcal{H}_{\rm C} \). The first term in \( F \) quantifies leakage, while the second term measures the deviation from the ideal operation.

Figure~\ref{fig: infidelity and leakage} illustrates the effects of deviations in \( J \) and \( V \) on the fidelity and leakage of the CZ gate. The \( x \)-axis represents variations in \( J \) (in units of \( J_{\text{CP}} \)), and the \( y \)-axis shows deviations in \( V \) (in units of \( |V_{\text{CP}}| \)). The left panels present a broad deviation range, highlighting uncontrollable behaviors for large parameter variations. The right panels focus on a narrower region around the optimal values \( J = J_{\text{CP}} \) and \( V = V_{\text{CP}} \), corresponding to realistic noise levels in superconducting circuits.

For a typical coupling strength \( J / 2\pi = 40 \, \text{MHz} \), deviations of \( \delta J / 2\pi =0.01J / 2\pi\approx 0.4 \, \text{MHz} \) and \(\delta V / 2\pi \approx 0.4 \, \text{MHz} \) well capture realistic experimental noise. Due to the large value of \( U \), rapid phase accumulation by the states \( \ket{00;20} \) and \( \ket{00;02} \) significantly contributes to the \( -\pi \) phase, while contributions from ZZ interaction is much weaker. Consequently, deviations in \( J \) have a greater impact on gate fidelity than those in \( V \). Leakage is even more sensitive to \( J \), as deviations hinder the return to the computational subspace. 

In conclusion, implementing the two-qubit CZ gate is feasible when considering the parameter deviations present in superconducting circuits.

\subsection{Detuning of Neighboring Transmons}
Superconducting qubits often exhibit inhomogeneities due to fabrication precision limitations. One significant source of errors in gate operations is detuning between adjacent qubits. This subsection explores the impact of detuning on the implementation of a CZ gate between the \( i \)-th and \( i+1 \)-th logical qubits, particularly focusing on the influence of detuning noise between the second rows of the two columns. The detuning is defined as:
\begin{equation}
\Delta \equiv \mu_{i+1,1} - \mu_{i,1}.
\end{equation}
This leads to a modified Hamiltonian:
\begin{equation}
\hat{H}_{\text{CP,detuned}} = \hat{H}_{\text{CP}} + \Delta \hat{n}_{i+1,1}.
\end{equation}
Due to the detuning, we expect imperfect evolution, resulting in leakage from the computational subspace. Specifically, the final state may have nonzero support in the orthogonal subspace, and we quantify the gate performance by examining both fidelity and leakage.

Given that the relevant subspaces are of dimension 2 or 3, we can analytically compute the average gate fidelity under detuning noise without additional assumptions. For simplicity, we assume a small detuning \( \Delta \), such that \( \Delta / J \ll 1 \), and investigate the fidelity dependence on \( \Delta \).

We begin by recalling the four subspaces from Eq.~\eqref{eq: four subspaces} and analyze each subspace in turn. The action of \( \hat{H}_{\text{CP,detuned}} \) in the space \( \mathcal{H}_1 \) remains unchanged, identical to that of \( \hat{H}_{\text{CP}} \).

In \( \mathcal{H}_2 \), the modified Hamiltonian is:
\begin{equation}
\hat{H}_2^{\prime} = \begin{pmatrix}
\Delta & -J \\
-J & 0
\end{pmatrix}.
\end{equation}
The eigenvalues are:
\begin{equation}
\lambda_{\pm} = \frac{\Delta \pm \sqrt{\Delta^2 + 4J^2}}{2},
\end{equation}
and the corresponding eigenstates are:
\begin{equation}
\ket{\phi_{\pm}} = \alpha_{\pm} \begin{pmatrix}
\lambda_{\pm}/J \\ -1
\end{pmatrix},
\end{equation}
where \( \alpha_{\pm} = 1/{\sqrt{1 + \lambda_{\pm}^2 / J^2}} \) are normalization coefficients. 
The state \( \ket{0;1}_L \) should be invariant under the perfect operation $\hat{U}_{\text{CZ}}$.
Initializing the state \( \ket{0;1}_L \) and evolving it under \( H_2^{\prime} \), the coefficient of \( \ket{0;1}_L \) at the gate time \( T_{\rm CP} \) becomes:
\begin{equation}
c_2 = \left(\frac{\alpha_{+} \lambda_{+}}{J}\right)^2 e^{-i \lambda_{+} T_{\rm CP}} + \left(\frac{\alpha_{-} \lambda_{-}}{J}\right)^2 e^{-i \lambda_{-} T_{\rm CP}}.
\end{equation}
Assuming a small \( \delta = \Delta / J \), we obtain \( c_2 \approx e^{-i \pi \delta} \).

In \( \mathcal{H}_3 \), the Hamiltonian is:
\begin{equation}
\hat{H}_3^{\prime} = \begin{pmatrix}
0 & -J \\
-J & \Delta
\end{pmatrix},
\end{equation}
which is related to \( \hat{H}_2^{\prime} \) by a basis permutation. The coefficient of the logical state \( \ket{10}_L \) is:
\begin{equation}
c_3 = \alpha_+^2 e^{-i \lambda_+ T_{\rm CP}} + \alpha_-^2 e^{-i \lambda_- T_{\rm CP}},
\end{equation}
which simplifies to \( c_3 \approx e^{-i \pi \delta} \).

Finally, in \( \mathcal{H}_4 \), the modified Hamiltonian is:
\begin{equation}
\hat{H}_4^{\prime} = \begin{pmatrix}
V + \Delta & -\sqrt{2} J & -\sqrt{2} J \\
-\sqrt{2} J & U & 0 \\
-\sqrt{2} J & 0 & U + 2 \Delta
\end{pmatrix}.
\end{equation}
Using the parameters \( V = k_1 J \) and \( U = k_2 J \), where \( k_1 = -\frac{7}{2} + 2\sqrt{3} \) and \( k_2 = -\left(\frac{7}{2} + 2\sqrt{3}\right) \), we express \( \hat{H}_4^{\prime} \) as:
\begin{equation}
\hat{H}_4^{\prime} = J(H_0 + \delta D).
\end{equation}
The perturbed eigenvalues are:
\[
\lambda_1 = (k_2 + \delta) J, \quad \lambda_2 = \left(-\frac{15}{2} + \delta\right) J, \quad \lambda_3 = \left(\frac{1}{2} + \delta\right) J.
\]
For a state initialized as \( \ket{\psi(0)} = (1, 0, 0)^{\mathrm{T}} \), we compute the coefficients \( \beta_i \equiv \braket{\phi_i}{\psi(0)} \), where \( \ket{\phi_i} \) are the eigenstates of \( \hat{H}_4^{\prime} \):
\begin{equation}
\beta_1 = \frac{1}{2} \delta, \quad \beta_2 = \frac{\sqrt{2} (\sqrt{3} - 1)}{4}, \quad \beta_3 = \frac{\sqrt{2} (-\sqrt{3} - 1)}{4}.
\end{equation}
The coefficient for \( \ket{11}_L \) at the gate time is:
\[
c_4 = \sum_{k=1,2,3} \beta_k^2 e^{-i \lambda_k T_{\rm CP}} \approx -\left( 1 + \frac{\delta^2}{4} e^{i 4 \sqrt{3} \pi} \right) e^{-2 \pi i \delta}.
\]

The projected matrix is \( P U P = \text{diag}(1, c_2, c_3, c_4) \). The fidelity \( F(\Delta) \) and leakage \( L(\Delta) \) are:
\[
F(\Delta) = 1 + \frac{\cos(4 \sqrt{3} \pi) \delta^2}{10} - \frac{2 \pi^2 \delta^2}{5}, \quad L(\Delta) = -\frac{\cos(4 \sqrt{3} \pi) \delta^2}{16}.
\]
The quadratic dependence of fidelity and leakage on \( \Delta \) demonstrates the robustness of the CZ gate against detuning noise. For a typical detuning of \( \Delta / 2\pi = 0.1 \) MHz and \( J / 2\pi = 40 \) MHz, we obtain \( \delta = 0.0025 \), leading to an infidelity of approximately \( 2.5 \times 10^{-5} \), confirming the applicability of the CZ gate under detuning noise.

\begin{figure}[t]
\centering
\includegraphics[width=0.8\columnwidth]{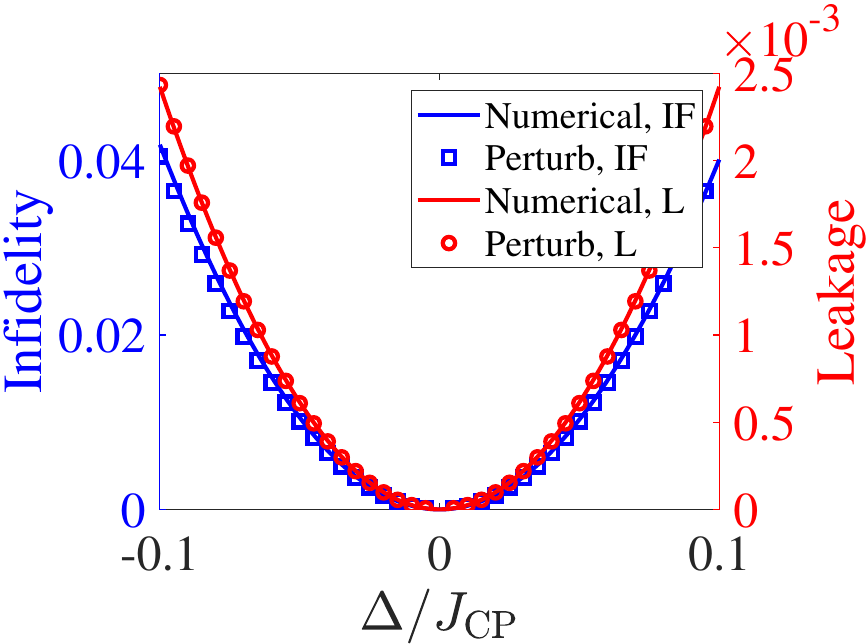}
    \caption{
    \textbf{Performance of the CZ gate under detuning noise.} The $x$-axis represents the detuning strength \( \Delta \), expressed in units of \( J_{\text{CP}} \). The left and right $y$-axes correspond to the CZ gate's infidelity and leakage, respectively. The solid lines depict the results from exact numerical simulations, while the squares and circles represent the second-order perturbation theory. The close agreement between the exact and perturbation results confirms a quadratic dependence of both infidelity and leakage on \( \Delta \).
    }
    \label{fig: detuning}
\end{figure}

Numerical results in Fig.~\ref{fig: detuning} validate the quadratic dependence of fidelity and leakage on \( \Delta \), in good agreement with the perturbation analysis.

\section{Circuit Example: Preparation of a GHZ State}

\begin{figure}[b]
	\centering
	\includegraphics[width=\columnwidth]{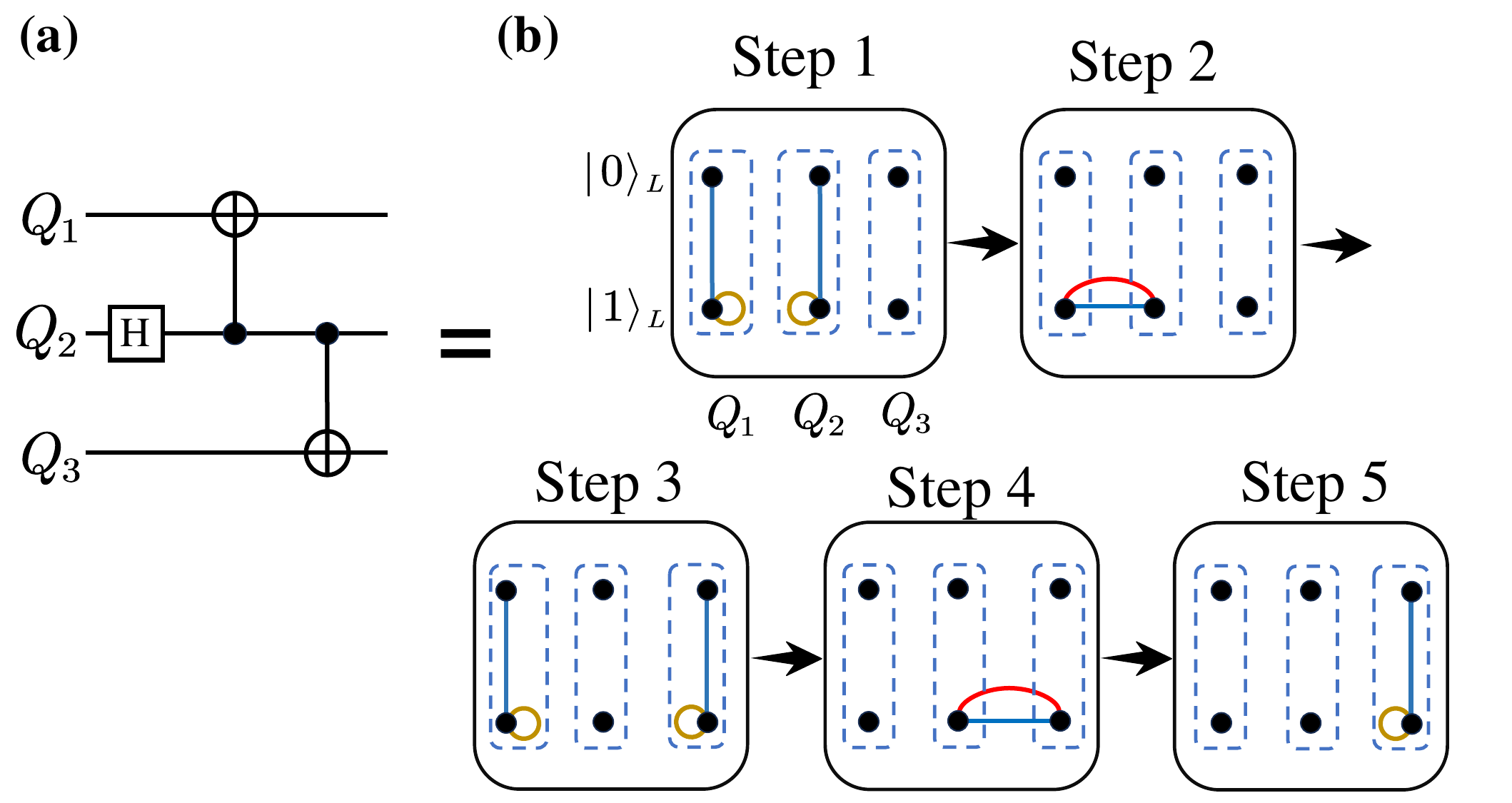}
	\caption{\textbf{GHZ state preparation.} (a) Logic circuit diagram for preparing a 3-qubit GHZ state, consisting of a Hadamard gate on \(Q_2\), followed by two CNOT gates on \(Q_1, Q_2\) and \(Q_2, Q_3\), respectively. (b) Hamiltonian graphs for preparing the GHZ state using dual-rail encoding. Steps 1 and 3 involve dual-role Hadamard gates, while steps 1 to 3 implement the CNOT gate on \(Q_1, Q_2\) and steps 3 to 5 implement the second CNOT gate on \(Q_2, Q_3\).}
	\label{fig: GHZ}
\end{figure}

The Greenberger-Horne-Zeilinger (GHZ) state is a highly entangled quantum state involving multiple qubits and is an essential example of multi-qubit entanglement. For three qubits, the GHZ state is expressed as
\begin{equation}
\ket{\text{GHZ}} = \frac{1}{\sqrt{2}}(\ket{000} + \ket{111}).
\end{equation}
The preparation and manipulation of GHZ states are important benchmarks for quantum hardware performance and are crucial for the development of scalable quantum technologies.

As shown in Fig.~\ref{fig: GHZ}, a 3-qubit GHZ state can be efficiently prepared using dual-rail encoded qubits. Fig.~\ref{fig: GHZ}(a) presents the logic circuit for generating the state, which involves a Hadamard gate on \(Q_2\), followed by two CNOT gates, one between \(Q_1\) and \(Q_2\), and another between \(Q_2\) and \(Q_3\). Fig.~\ref{fig: GHZ}(b) shows the corresponding quantum walk graphs at each step. The preparation process requires two Hadamard gates and one CZ gate to implement the CNOT gate. Notably, single-qubit gates can be applied simultaneously on different logic qubits. In total, the GHZ state can be prepared in five steps for dual-rail encoded qubits: the first step involves a Hadamard gate on \(Q_2\), followed by the implementation of the CNOT gate between \(Q_1\) and \(Q_2\) from steps 1 to 3, and the final CNOT gate between \(Q_2\) and \(Q_3\) from steps 3 to 5.

\section{Discussion}

The following key points merit further discussion regarding this framework:

\begin{itemize}
    \item Bosons hopping on a lattice described by the Bose-Hubbard model can be interpreted as multiple quantum walkers on a graph whose vertices and edges encode the sites and couplings, respectively. Multiple walkers on $2\times n$ vertices provide exponentially growing resources to accommodate the dual-rail encoding scheme.
    \item After adding a small ZZ interaction, the extended Bose-Hubbard model can be utilized to generate maximum entangling logic gates and a universal set of gates. These logic gates operate on dual-rail encoded qubits, each comprising a pair of resonantly coupled transmons and an additional detuned coupler. This system enables perfect adiabatic control without requiring any pulses. It leverages the ability to access the non-computational subspace during evolution while being entirely confined to the computational subspace by the end of the evolution. The gates remain unitary when the Hamiltonian parameters are appropriately tuned.
    \item On the 2-dimensional dual-rail encoded qubit array, we propose schemes for implementing a two-qubit CPhase gate, an iSWAP gate, and a three-qubit CCPhase gate in the transverse connection, along with a CPhase gate in the longitudinal connection. Practical parameters are chosen for the tunable coupling scheme of the superconducting circuit. Specifically, the two-qubit gates, CPhase and iSWAP, are flawlessly constructed with fidelity equal to 1. In contrast, the CCPhase gate is constrained to discrete phases up to \( -0.271\pi \) when fidelity \( \geqslant 0.99 \). We numerically verify the small impact of dephasing, relaxation noises, imperfect coupling and ZZ interaction strengths on the performance of the CZ gate. We also analytically derive that the infidelity and leakage error are first-order insensitivity to detuning between two neighboring physical qubits for the CZ gate.
    \item As experimentally demonstrated with transmons, dual-rail encoded qubits exhibit high coherence, allowing leakage errors to be converted into erasure errors with more favorable thresholds for error correction~\cite{levine2024demonstrating}. This framework also allows constructing dual-rail qubits from other physical systems with long \( T_1 \) but relatively shorter \( T_2 \) coherence times.
    \item Our proposed architecture is hardware-efficient because it requires fewer hardware modifications, lower resource costs, and straightforward integration with today’s superconducting qubit technologies. Specifically: 1. It aligns with existing transmon and tunable coupler designs. Dual-rail encoding can be realized by pairs of transmons and a flux-tunable coupler—both of which are standard elements in many superconducting quantum devices. 2. It avoids reliance on large numbers of ancillary qubits or intricate multi-level control schemes. By encoding a single photon-number excitation across two transmons and employing CTQW, the architecture stays relatively simple and utilizes hardware components already typical in current experiments. 3. It naturally converts leakage and relaxation into erasure events, simplifying error detection and correction rather than requiring extra measurement circuitry or feedback loops. 
    \item Although our analysis and examples focus on superconducting transmons, the concept of implementing CTQW on dual-rail encoded qubits can be extended to a variety of other physical systems. Any platform supporting sufficiently coherent bosonic excitations and controllable interactions---such as neutral atoms in optical lattices, trapped ions, or photonic waveguide arrays---is amenable to the proposed scheme. 
\end{itemize}

\section{Conclusion}

In this work, we have demonstrated how combining dual-rail qubit encoding with continuous-time quantum walks (CTQW) in superconducting architectures offers a hardware-efficient route toward robust and scalable quantum computation. We harnessed photon-number-conserving dynamics to suppress population leakage and enable high-fidelity single-, two-, and three-qubit gate operations, thereby matching the goal of dual-rail encoded qubits for the conversion of erasure errors. Our theoretical and numerical results indicate that this synergy not only mitigates conventional sources of decoherence (e.g., dephasing and relaxation) but also remains tolerant to coupling imperfections, underscoring the resilience of the dual-rail framework.

Beyond the immediate gains in gate fidelity and error correction, this approach can serve as a foundation for early fault-tolerant quantum computing. Since erasure errors are more tractable in quantum error-correcting codes, adopting a dual-rail strategy may simplify the overhead associated with representative fault-tolerant protocols. Moreover, by exploiting tunable coupler strengths—already feasible in present-day superconducting devices—our method aligns well with current experimental capabilities, paving the way for near-term demonstrations of larger-scale quantum systems.

\acknowledgments
The manuscript was collaboratively authored by all contributors. HY Guan and Y Li made equal contributions to the derivation, the execution of numerical simulations, and the plotting of figures. The project was initiated, overseen, and coordinated by XH Deng. This work was supported by the Key-Area Research and Development Program of Guang-Dong Province (Grant No. 2018B030326001), and the Science, Technology and Innovation Commission of Shenzhen Municipality (JCYJ20170412152620376, KYTDPT20181011104202253), and the Shenzhen Science and Technology Program (KQTD20200820113010023).


\begin{thebibliography}{48}%
\makeatletter
\providecommand \@ifxundefined [1]{%
 \@ifx{#1\undefined}
}%
\providecommand \@ifnum [1]{%
 \ifnum #1\expandafter \@firstoftwo
 \else \expandafter \@secondoftwo
 \fi
}%
\providecommand \@ifx [1]{%
 \ifx #1\expandafter \@firstoftwo
 \else \expandafter \@secondoftwo
 \fi
}%
\providecommand \natexlab [1]{#1}%
\providecommand \enquote  [1]{``#1''}%
\providecommand \bibnamefont  [1]{#1}%
\providecommand \bibfnamefont [1]{#1}%
\providecommand \citenamefont [1]{#1}%
\providecommand \href@noop [0]{\@secondoftwo}%
\providecommand \href [0]{\begingroup \@sanitize@url \@href}%
\providecommand \@href[1]{\@@startlink{#1}\@@href}%
\providecommand \@@href[1]{\endgroup#1\@@endlink}%
\providecommand \@sanitize@url [0]{\catcode `\\12\catcode `\$12\catcode
  `\&12\catcode `\#12\catcode `\^12\catcode `\_12\catcode `\%12\relax}%
\providecommand \@@startlink[1]{}%
\providecommand \@@endlink[0]{}%
\providecommand \url  [0]{\begingroup\@sanitize@url \@url }%
\providecommand \@url [1]{\endgroup\@href {#1}{\urlprefix }}%
\providecommand \urlprefix  [0]{URL }%
\providecommand \Eprint [0]{\href }%
\providecommand \doibase [0]{https://doi.org/}%
\providecommand \selectlanguage [0]{\@gobble}%
\providecommand \bibinfo  [0]{\@secondoftwo}%
\providecommand \bibfield  [0]{\@secondoftwo}%
\providecommand \translation [1]{[#1]}%
\providecommand \BibitemOpen [0]{}%
\providecommand \bibitemStop [0]{}%
\providecommand \bibitemNoStop [0]{.\EOS\space}%
\providecommand \EOS [0]{\spacefactor3000\relax}%
\providecommand \BibitemShut  [1]{\csname bibitem#1\endcsname}%
\let\auto@bib@innerbib\@empty
\bibitem [{\citenamefont {Kubica}\ \emph {et~al.}(2023)\citenamefont {Kubica},
  \citenamefont {Haim}, \citenamefont {Vaknin}, \citenamefont {Levine},
  \citenamefont {Brand{\~a}o},\ and\ \citenamefont
  {Retzker}}]{kubica2023erasure}%
  \BibitemOpen
  \bibfield  {author} {\bibinfo {author} {\bibfnamefont {A.}~\bibnamefont
  {Kubica}}, \bibinfo {author} {\bibfnamefont {A.}~\bibnamefont {Haim}},
  \bibinfo {author} {\bibfnamefont {Y.}~\bibnamefont {Vaknin}}, \bibinfo
  {author} {\bibfnamefont {H.}~\bibnamefont {Levine}}, \bibinfo {author}
  {\bibfnamefont {F.}~\bibnamefont {Brand{\~a}o}},\ and\ \bibinfo {author}
  {\bibfnamefont {A.}~\bibnamefont {Retzker}},\ }\bibfield  {title} {\bibinfo
  {title} {Erasure qubits: Overcoming the $t_1$ limit in superconducting
  circuits},\ }\href@noop {} {\bibfield  {journal} {\bibinfo  {journal}
  {Physical Review X}\ }\textbf {\bibinfo {volume} {13}},\ \bibinfo {pages}
  {041022} (\bibinfo {year} {2023})}\BibitemShut {NoStop}%
\bibitem [{\citenamefont {Shim}\ and\ \citenamefont
  {Tahan}(2016)}]{shim2016semiconductor}%
  \BibitemOpen
  \bibfield  {author} {\bibinfo {author} {\bibfnamefont {Y.-P.}\ \bibnamefont
  {Shim}}\ and\ \bibinfo {author} {\bibfnamefont {C.}~\bibnamefont {Tahan}},\
  }\bibfield  {title} {\bibinfo {title} {Semiconductor-inspired design
  principles for superconducting quantum computing},\ }\href@noop {} {\bibfield
   {journal} {\bibinfo  {journal} {Nature communications}\ }\textbf {\bibinfo
  {volume} {7}},\ \bibinfo {pages} {11059} (\bibinfo {year}
  {2016})}\BibitemShut {NoStop}%
\bibitem [{\citenamefont {Teoh}\ \emph {et~al.}(2023)\citenamefont {Teoh},
  \citenamefont {Winkel}, \citenamefont {Babla}, \citenamefont {Chapman},
  \citenamefont {Claes}, \citenamefont {de~Graaf}, \citenamefont {Garmon},
  \citenamefont {Kalfus}, \citenamefont {Lu}, \citenamefont {Maiti} \emph
  {et~al.}}]{teoh2023dual}%
  \BibitemOpen
  \bibfield  {author} {\bibinfo {author} {\bibfnamefont {J.~D.}\ \bibnamefont
  {Teoh}}, \bibinfo {author} {\bibfnamefont {P.}~\bibnamefont {Winkel}},
  \bibinfo {author} {\bibfnamefont {H.~K.}\ \bibnamefont {Babla}}, \bibinfo
  {author} {\bibfnamefont {B.~J.}\ \bibnamefont {Chapman}}, \bibinfo {author}
  {\bibfnamefont {J.}~\bibnamefont {Claes}}, \bibinfo {author} {\bibfnamefont
  {S.~J.}\ \bibnamefont {de~Graaf}}, \bibinfo {author} {\bibfnamefont {J.~W.}\
  \bibnamefont {Garmon}}, \bibinfo {author} {\bibfnamefont {W.~D.}\
  \bibnamefont {Kalfus}}, \bibinfo {author} {\bibfnamefont {Y.}~\bibnamefont
  {Lu}}, \bibinfo {author} {\bibfnamefont {A.}~\bibnamefont {Maiti}}, \emph
  {et~al.},\ }\bibfield  {title} {\bibinfo {title} {Dual-rail encoding with
  superconducting cavities},\ }\href@noop {} {\bibfield  {journal} {\bibinfo
  {journal} {Proceedings of the National Academy of Sciences}\ }\textbf
  {\bibinfo {volume} {120}},\ \bibinfo {pages} {e2221736120} (\bibinfo {year}
  {2023})}\BibitemShut {NoStop}%
\bibitem [{\citenamefont {Campbell}\ \emph {et~al.}(2020)\citenamefont
  {Campbell}, \citenamefont {Shim}, \citenamefont {Kannan}, \citenamefont
  {Winik}, \citenamefont {Kim}, \citenamefont {Melville}, \citenamefont
  {Niedzielski}, \citenamefont {Yoder}, \citenamefont {Tahan}, \citenamefont
  {Gustavsson} \emph {et~al.}}]{campbell2020universal}%
  \BibitemOpen
  \bibfield  {author} {\bibinfo {author} {\bibfnamefont {D.~L.}\ \bibnamefont
  {Campbell}}, \bibinfo {author} {\bibfnamefont {Y.-P.}\ \bibnamefont {Shim}},
  \bibinfo {author} {\bibfnamefont {B.}~\bibnamefont {Kannan}}, \bibinfo
  {author} {\bibfnamefont {R.}~\bibnamefont {Winik}}, \bibinfo {author}
  {\bibfnamefont {D.~K.}\ \bibnamefont {Kim}}, \bibinfo {author} {\bibfnamefont
  {A.}~\bibnamefont {Melville}}, \bibinfo {author} {\bibfnamefont {B.~M.}\
  \bibnamefont {Niedzielski}}, \bibinfo {author} {\bibfnamefont {J.~L.}\
  \bibnamefont {Yoder}}, \bibinfo {author} {\bibfnamefont {C.}~\bibnamefont
  {Tahan}}, \bibinfo {author} {\bibfnamefont {S.}~\bibnamefont {Gustavsson}},
  \emph {et~al.},\ }\bibfield  {title} {\bibinfo {title} {Universal
  nonadiabatic control of small-gap superconducting qubits},\ }\href@noop {}
  {\bibfield  {journal} {\bibinfo  {journal} {Physical Review X}\ }\textbf
  {\bibinfo {volume} {10}},\ \bibinfo {pages} {041051} (\bibinfo {year}
  {2020})}\BibitemShut {NoStop}%
\bibitem [{\citenamefont {Levine}\ \emph {et~al.}(2024)\citenamefont {Levine},
  \citenamefont {Haim}, \citenamefont {Hung}, \citenamefont {Alidoust},
  \citenamefont {Kalaee}, \citenamefont {DeLorenzo}, \citenamefont {Wollack},
  \citenamefont {Arrangoiz-Arriola}, \citenamefont {Khalajhedayati},
  \citenamefont {Sanil} \emph {et~al.}}]{levine2024demonstrating}%
  \BibitemOpen
  \bibfield  {author} {\bibinfo {author} {\bibfnamefont {H.}~\bibnamefont
  {Levine}}, \bibinfo {author} {\bibfnamefont {A.}~\bibnamefont {Haim}},
  \bibinfo {author} {\bibfnamefont {J.~S.}\ \bibnamefont {Hung}}, \bibinfo
  {author} {\bibfnamefont {N.}~\bibnamefont {Alidoust}}, \bibinfo {author}
  {\bibfnamefont {M.}~\bibnamefont {Kalaee}}, \bibinfo {author} {\bibfnamefont
  {L.}~\bibnamefont {DeLorenzo}}, \bibinfo {author} {\bibfnamefont {E.~A.}\
  \bibnamefont {Wollack}}, \bibinfo {author} {\bibfnamefont {P.}~\bibnamefont
  {Arrangoiz-Arriola}}, \bibinfo {author} {\bibfnamefont {A.}~\bibnamefont
  {Khalajhedayati}}, \bibinfo {author} {\bibfnamefont {R.}~\bibnamefont
  {Sanil}}, \emph {et~al.},\ }\bibfield  {title} {\bibinfo {title}
  {Demonstrating a long-coherence dual-rail erasure qubit using tunable
  transmons},\ }\href@noop {} {\bibfield  {journal} {\bibinfo  {journal}
  {Physical Review X}\ }\textbf {\bibinfo {volume} {14}},\ \bibinfo {pages}
  {011051} (\bibinfo {year} {2024})}\BibitemShut {NoStop}%
\bibitem [{\citenamefont {Chou}\ \emph {et~al.}(2024)\citenamefont {Chou},
  \citenamefont {Shemma}, \citenamefont {McCarrick}, \citenamefont {Chien},
  \citenamefont {Teoh}, \citenamefont {Winkel}, \citenamefont {Anderson},
  \citenamefont {Chen}, \citenamefont {Curtis}, \citenamefont {de~Graaf} \emph
  {et~al.}}]{chou2024superconducting}%
  \BibitemOpen
  \bibfield  {author} {\bibinfo {author} {\bibfnamefont {K.~S.}\ \bibnamefont
  {Chou}}, \bibinfo {author} {\bibfnamefont {T.}~\bibnamefont {Shemma}},
  \bibinfo {author} {\bibfnamefont {H.}~\bibnamefont {McCarrick}}, \bibinfo
  {author} {\bibfnamefont {T.-C.}\ \bibnamefont {Chien}}, \bibinfo {author}
  {\bibfnamefont {J.~D.}\ \bibnamefont {Teoh}}, \bibinfo {author}
  {\bibfnamefont {P.}~\bibnamefont {Winkel}}, \bibinfo {author} {\bibfnamefont
  {A.}~\bibnamefont {Anderson}}, \bibinfo {author} {\bibfnamefont
  {J.}~\bibnamefont {Chen}}, \bibinfo {author} {\bibfnamefont {J.~C.}\
  \bibnamefont {Curtis}}, \bibinfo {author} {\bibfnamefont {S.~J.}\
  \bibnamefont {de~Graaf}}, \emph {et~al.},\ }\bibfield  {title} {\bibinfo
  {title} {A superconducting dual-rail cavity qubit with erasure-detected
  logical measurements},\ }\href@noop {} {\bibfield  {journal} {\bibinfo
  {journal} {Nature Physics}\ }\textbf {\bibinfo {volume} {20}},\ \bibinfo
  {pages} {1454} (\bibinfo {year} {2024})}\BibitemShut {NoStop}%
\bibitem [{\citenamefont {Koottandavida}\ \emph {et~al.}(2024)\citenamefont
  {Koottandavida}, \citenamefont {Tsioutsios}, \citenamefont {Kargioti},
  \citenamefont {Smith}, \citenamefont {Joshi}, \citenamefont {Dai},
  \citenamefont {Teoh}, \citenamefont {Curtis}, \citenamefont {Frunzio},
  \citenamefont {Schoelkopf} \emph {et~al.}}]{koottandavida2024erasure}%
  \BibitemOpen
  \bibfield  {author} {\bibinfo {author} {\bibfnamefont {A.}~\bibnamefont
  {Koottandavida}}, \bibinfo {author} {\bibfnamefont {I.}~\bibnamefont
  {Tsioutsios}}, \bibinfo {author} {\bibfnamefont {A.}~\bibnamefont
  {Kargioti}}, \bibinfo {author} {\bibfnamefont {C.~R.}\ \bibnamefont {Smith}},
  \bibinfo {author} {\bibfnamefont {V.~R.}\ \bibnamefont {Joshi}}, \bibinfo
  {author} {\bibfnamefont {W.}~\bibnamefont {Dai}}, \bibinfo {author}
  {\bibfnamefont {J.~D.}\ \bibnamefont {Teoh}}, \bibinfo {author}
  {\bibfnamefont {J.~C.}\ \bibnamefont {Curtis}}, \bibinfo {author}
  {\bibfnamefont {L.}~\bibnamefont {Frunzio}}, \bibinfo {author} {\bibfnamefont
  {R.~J.}\ \bibnamefont {Schoelkopf}}, \emph {et~al.},\ }\bibfield  {title}
  {\bibinfo {title} {Erasure detection of a dual-rail qubit encoded in a
  double-post superconducting cavity},\ }\href@noop {} {\bibfield  {journal}
  {\bibinfo  {journal} {Physical Review Letters}\ }\textbf {\bibinfo {volume}
  {132}},\ \bibinfo {pages} {180601} (\bibinfo {year} {2024})}\BibitemShut
  {NoStop}%
\bibitem [{\citenamefont {Weiss}\ \emph {et~al.}(2024)\citenamefont {Weiss},
  \citenamefont {Puri},\ and\ \citenamefont {Girvin}}]{weiss2024quantum}%
  \BibitemOpen
  \bibfield  {author} {\bibinfo {author} {\bibfnamefont {D.}~\bibnamefont
  {Weiss}}, \bibinfo {author} {\bibfnamefont {S.}~\bibnamefont {Puri}},\ and\
  \bibinfo {author} {\bibfnamefont {S.}~\bibnamefont {Girvin}},\ }\bibfield
  {title} {\bibinfo {title} {Quantum random access memory architectures using
  3d superconducting cavities},\ }\href@noop {} {\bibfield  {journal} {\bibinfo
   {journal} {PRX Quantum}\ }\textbf {\bibinfo {volume} {5}},\ \bibinfo {pages}
  {020312} (\bibinfo {year} {2024})}\BibitemShut {NoStop}%
\bibitem [{\citenamefont {Wong}(2019)}]{wong2019isolated}%
  \BibitemOpen
  \bibfield  {author} {\bibinfo {author} {\bibfnamefont {T.~G.}\ \bibnamefont
  {Wong}},\ }\bibfield  {title} {\bibinfo {title} {Isolated vertices in
  continuous-time quantum walks on dynamic graphs},\ }\href@noop {} {\bibfield
  {journal} {\bibinfo  {journal} {Physical Review A}\ }\textbf {\bibinfo
  {volume} {100}},\ \bibinfo {pages} {062325} (\bibinfo {year}
  {2019})}\BibitemShut {NoStop}%
\bibitem [{\citenamefont {Chawla}\ \emph {et~al.}(2023)\citenamefont {Chawla},
  \citenamefont {Singh}, \citenamefont {Agarwal}, \citenamefont {Srinivasan},\
  and\ \citenamefont {Chandrashekar}}]{chawla2023multi}%
  \BibitemOpen
  \bibfield  {author} {\bibinfo {author} {\bibfnamefont {P.}~\bibnamefont
  {Chawla}}, \bibinfo {author} {\bibfnamefont {S.}~\bibnamefont {Singh}},
  \bibinfo {author} {\bibfnamefont {A.}~\bibnamefont {Agarwal}}, \bibinfo
  {author} {\bibfnamefont {S.}~\bibnamefont {Srinivasan}},\ and\ \bibinfo
  {author} {\bibfnamefont {C.}~\bibnamefont {Chandrashekar}},\ }\bibfield
  {title} {\bibinfo {title} {Multi-qubit quantum computing using discrete-time
  quantum walks on closed graphs},\ }\href@noop {} {\bibfield  {journal}
  {\bibinfo  {journal} {Scientific Reports}\ }\textbf {\bibinfo {volume}
  {13}},\ \bibinfo {pages} {12078} (\bibinfo {year} {2023})}\BibitemShut
  {NoStop}%
\bibitem [{\citenamefont {Farhi}\ and\ \citenamefont
  {Gutmann}(1998)}]{farhi1998quantum}%
  \BibitemOpen
  \bibfield  {author} {\bibinfo {author} {\bibfnamefont {E.}~\bibnamefont
  {Farhi}}\ and\ \bibinfo {author} {\bibfnamefont {S.}~\bibnamefont
  {Gutmann}},\ }\bibfield  {title} {\bibinfo {title} {Quantum computation and
  decision trees},\ }\href@noop {} {\bibfield  {journal} {\bibinfo  {journal}
  {Physical Review A}\ }\textbf {\bibinfo {volume} {58}},\ \bibinfo {pages}
  {915} (\bibinfo {year} {1998})}\BibitemShut {NoStop}%
\bibitem [{\citenamefont {Childs}\ \emph {et~al.}(2002)\citenamefont {Childs},
  \citenamefont {Farhi},\ and\ \citenamefont {Gutmann}}]{childs2002example}%
  \BibitemOpen
  \bibfield  {author} {\bibinfo {author} {\bibfnamefont {A.~M.}\ \bibnamefont
  {Childs}}, \bibinfo {author} {\bibfnamefont {E.}~\bibnamefont {Farhi}},\ and\
  \bibinfo {author} {\bibfnamefont {S.}~\bibnamefont {Gutmann}},\ }\bibfield
  {title} {\bibinfo {title} {An example of the difference between quantum and
  classical random walks},\ }\href@noop {} {\bibfield  {journal} {\bibinfo
  {journal} {Quantum Information Processing}\ }\textbf {\bibinfo {volume}
  {1}},\ \bibinfo {pages} {35} (\bibinfo {year} {2002})}\BibitemShut {NoStop}%
\bibitem [{\citenamefont {Childs}\ \emph {et~al.}(2003)\citenamefont {Childs},
  \citenamefont {Cleve}, \citenamefont {Deotto}, \citenamefont {Farhi},
  \citenamefont {Gutmann},\ and\ \citenamefont
  {Spielman}}]{childs2003exponential}%
  \BibitemOpen
  \bibfield  {author} {\bibinfo {author} {\bibfnamefont {A.~M.}\ \bibnamefont
  {Childs}}, \bibinfo {author} {\bibfnamefont {R.}~\bibnamefont {Cleve}},
  \bibinfo {author} {\bibfnamefont {E.}~\bibnamefont {Deotto}}, \bibinfo
  {author} {\bibfnamefont {E.}~\bibnamefont {Farhi}}, \bibinfo {author}
  {\bibfnamefont {S.}~\bibnamefont {Gutmann}},\ and\ \bibinfo {author}
  {\bibfnamefont {D.~A.}\ \bibnamefont {Spielman}},\ }\bibfield  {title}
  {\bibinfo {title} {Exponential algorithmic speedup by a quantum walk},\ }in\
  \href@noop {} {\emph {\bibinfo {booktitle} {Proceedings of the thirty-fifth
  annual ACM symposium on Theory of computing}}}\ (\bibinfo {year} {2003})\
  pp.\ \bibinfo {pages} {59--68}\BibitemShut {NoStop}%
\bibitem [{\citenamefont {Childs}(2004)}]{childs2004quantum}%
  \BibitemOpen
  \bibfield  {author} {\bibinfo {author} {\bibfnamefont {A.~M.}\ \bibnamefont
  {Childs}},\ }\emph {\bibinfo {title} {Quantum information processing in
  continuous time}},\ \href@noop {} {Ph.D. thesis},\ \bibinfo  {school}
  {Massachusetts Institute of Technology} (\bibinfo {year} {2004})\BibitemShut
  {NoStop}%
\bibitem [{\citenamefont {Peruzzo}\ \emph {et~al.}(2010)\citenamefont
  {Peruzzo}, \citenamefont {Lobino}, \citenamefont {Matthews}, \citenamefont
  {Matsuda}, \citenamefont {Politi}, \citenamefont {Poulios}, \citenamefont
  {Zhou}, \citenamefont {Lahini}, \citenamefont {Ismail}, \citenamefont
  {W{\"o}rhoff} \emph {et~al.}}]{peruzzo2010quantum}%
  \BibitemOpen
  \bibfield  {author} {\bibinfo {author} {\bibfnamefont {A.}~\bibnamefont
  {Peruzzo}}, \bibinfo {author} {\bibfnamefont {M.}~\bibnamefont {Lobino}},
  \bibinfo {author} {\bibfnamefont {J.~C.}\ \bibnamefont {Matthews}}, \bibinfo
  {author} {\bibfnamefont {N.}~\bibnamefont {Matsuda}}, \bibinfo {author}
  {\bibfnamefont {A.}~\bibnamefont {Politi}}, \bibinfo {author} {\bibfnamefont
  {K.}~\bibnamefont {Poulios}}, \bibinfo {author} {\bibfnamefont {X.-Q.}\
  \bibnamefont {Zhou}}, \bibinfo {author} {\bibfnamefont {Y.}~\bibnamefont
  {Lahini}}, \bibinfo {author} {\bibfnamefont {N.}~\bibnamefont {Ismail}},
  \bibinfo {author} {\bibfnamefont {K.}~\bibnamefont {W{\"o}rhoff}}, \emph
  {et~al.},\ }\bibfield  {title} {\bibinfo {title} {Quantum walks of correlated
  photons},\ }\href@noop {} {\bibfield  {journal} {\bibinfo  {journal}
  {Science}\ }\textbf {\bibinfo {volume} {329}},\ \bibinfo {pages} {1500}
  (\bibinfo {year} {2010})}\BibitemShut {NoStop}%
\bibitem [{\citenamefont {Yan}\ \emph {et~al.}(2019)\citenamefont {Yan},
  \citenamefont {Zhang}, \citenamefont {Gong}, \citenamefont {Wu},
  \citenamefont {Zheng}, \citenamefont {Li}, \citenamefont {Wang},
  \citenamefont {Liang}, \citenamefont {Lin}, \citenamefont {Xu} \emph
  {et~al.}}]{yan2019strongly}%
  \BibitemOpen
  \bibfield  {author} {\bibinfo {author} {\bibfnamefont {Z.}~\bibnamefont
  {Yan}}, \bibinfo {author} {\bibfnamefont {Y.-R.}\ \bibnamefont {Zhang}},
  \bibinfo {author} {\bibfnamefont {M.}~\bibnamefont {Gong}}, \bibinfo {author}
  {\bibfnamefont {Y.}~\bibnamefont {Wu}}, \bibinfo {author} {\bibfnamefont
  {Y.}~\bibnamefont {Zheng}}, \bibinfo {author} {\bibfnamefont
  {S.}~\bibnamefont {Li}}, \bibinfo {author} {\bibfnamefont {C.}~\bibnamefont
  {Wang}}, \bibinfo {author} {\bibfnamefont {F.}~\bibnamefont {Liang}},
  \bibinfo {author} {\bibfnamefont {J.}~\bibnamefont {Lin}}, \bibinfo {author}
  {\bibfnamefont {Y.}~\bibnamefont {Xu}}, \emph {et~al.},\ }\bibfield  {title}
  {\bibinfo {title} {Strongly correlated quantum walks with a 12-qubit
  superconducting processor},\ }\href@noop {} {\bibfield  {journal} {\bibinfo
  {journal} {Science}\ }\textbf {\bibinfo {volume} {364}},\ \bibinfo {pages}
  {753} (\bibinfo {year} {2019})}\BibitemShut {NoStop}%
\bibitem [{\citenamefont {Gong}\ \emph {et~al.}(2021)\citenamefont {Gong},
  \citenamefont {Wang}, \citenamefont {Zha}, \citenamefont {Chen},
  \citenamefont {Huang}, \citenamefont {Wu}, \citenamefont {Zhu}, \citenamefont
  {Zhao}, \citenamefont {Li}, \citenamefont {Guo} \emph
  {et~al.}}]{gong2021quantum}%
  \BibitemOpen
  \bibfield  {author} {\bibinfo {author} {\bibfnamefont {M.}~\bibnamefont
  {Gong}}, \bibinfo {author} {\bibfnamefont {S.}~\bibnamefont {Wang}}, \bibinfo
  {author} {\bibfnamefont {C.}~\bibnamefont {Zha}}, \bibinfo {author}
  {\bibfnamefont {M.-C.}\ \bibnamefont {Chen}}, \bibinfo {author}
  {\bibfnamefont {H.-L.}\ \bibnamefont {Huang}}, \bibinfo {author}
  {\bibfnamefont {Y.}~\bibnamefont {Wu}}, \bibinfo {author} {\bibfnamefont
  {Q.}~\bibnamefont {Zhu}}, \bibinfo {author} {\bibfnamefont {Y.}~\bibnamefont
  {Zhao}}, \bibinfo {author} {\bibfnamefont {S.}~\bibnamefont {Li}}, \bibinfo
  {author} {\bibfnamefont {S.}~\bibnamefont {Guo}}, \emph {et~al.},\ }\bibfield
   {title} {\bibinfo {title} {Quantum walks on a programmable two-dimensional
  62-qubit superconducting processor},\ }\href@noop {} {\bibfield  {journal}
  {\bibinfo  {journal} {Science}\ }\textbf {\bibinfo {volume} {372}},\ \bibinfo
  {pages} {948} (\bibinfo {year} {2021})}\BibitemShut {NoStop}%
\bibitem [{\citenamefont {Lahini}\ \emph {et~al.}(2012)\citenamefont {Lahini},
  \citenamefont {Verbin}, \citenamefont {Huber}, \citenamefont {Bromberg},
  \citenamefont {Pugatch},\ and\ \citenamefont
  {Silberberg}}]{lahini2012quantum}%
  \BibitemOpen
  \bibfield  {author} {\bibinfo {author} {\bibfnamefont {Y.}~\bibnamefont
  {Lahini}}, \bibinfo {author} {\bibfnamefont {M.}~\bibnamefont {Verbin}},
  \bibinfo {author} {\bibfnamefont {S.~D.}\ \bibnamefont {Huber}}, \bibinfo
  {author} {\bibfnamefont {Y.}~\bibnamefont {Bromberg}}, \bibinfo {author}
  {\bibfnamefont {R.}~\bibnamefont {Pugatch}},\ and\ \bibinfo {author}
  {\bibfnamefont {Y.}~\bibnamefont {Silberberg}},\ }\bibfield  {title}
  {\bibinfo {title} {Quantum walk of two interacting bosons},\ }\href@noop {}
  {\bibfield  {journal} {\bibinfo  {journal} {Physical Review A—Atomic,
  Molecular, and Optical Physics}\ }\textbf {\bibinfo {volume} {86}},\ \bibinfo
  {pages} {011603} (\bibinfo {year} {2012})}\BibitemShut {NoStop}%
\bibitem [{\citenamefont {Siloi}\ \emph {et~al.}(2017)\citenamefont {Siloi},
  \citenamefont {Benedetti}, \citenamefont {Piccinini}, \citenamefont {Piilo},
  \citenamefont {Maniscalco}, \citenamefont {Paris},\ and\ \citenamefont
  {Bordone}}]{siloi2017noisy}%
  \BibitemOpen
  \bibfield  {author} {\bibinfo {author} {\bibfnamefont {I.}~\bibnamefont
  {Siloi}}, \bibinfo {author} {\bibfnamefont {C.}~\bibnamefont {Benedetti}},
  \bibinfo {author} {\bibfnamefont {E.}~\bibnamefont {Piccinini}}, \bibinfo
  {author} {\bibfnamefont {J.}~\bibnamefont {Piilo}}, \bibinfo {author}
  {\bibfnamefont {S.}~\bibnamefont {Maniscalco}}, \bibinfo {author}
  {\bibfnamefont {M.~G.}\ \bibnamefont {Paris}},\ and\ \bibinfo {author}
  {\bibfnamefont {P.}~\bibnamefont {Bordone}},\ }\bibfield  {title} {\bibinfo
  {title} {Noisy quantum walks of two indistinguishable interacting
  particles},\ }\href@noop {} {\bibfield  {journal} {\bibinfo  {journal}
  {Physical Review A}\ }\textbf {\bibinfo {volume} {95}},\ \bibinfo {pages}
  {022106} (\bibinfo {year} {2017})}\BibitemShut {NoStop}%
\bibitem [{\citenamefont {Lewis}\ \emph {et~al.}(2021)\citenamefont {Lewis},
  \citenamefont {Benhemou}, \citenamefont {Feinstein}, \citenamefont {Banchi},\
  and\ \citenamefont {Bose}}]{lewis2021optimal}%
  \BibitemOpen
  \bibfield  {author} {\bibinfo {author} {\bibfnamefont {D.}~\bibnamefont
  {Lewis}}, \bibinfo {author} {\bibfnamefont {A.}~\bibnamefont {Benhemou}},
  \bibinfo {author} {\bibfnamefont {N.}~\bibnamefont {Feinstein}}, \bibinfo
  {author} {\bibfnamefont {L.}~\bibnamefont {Banchi}},\ and\ \bibinfo {author}
  {\bibfnamefont {S.}~\bibnamefont {Bose}},\ }\bibfield  {title} {\bibinfo
  {title} {Optimal quantum spatial search with one-dimensional long-range
  interactions},\ }\href@noop {} {\bibfield  {journal} {\bibinfo  {journal}
  {Physical Review Letters}\ }\textbf {\bibinfo {volume} {126}},\ \bibinfo
  {pages} {240502} (\bibinfo {year} {2021})}\BibitemShut {NoStop}%
\bibitem [{\citenamefont {Xing}\ \emph {et~al.}(2024)\citenamefont {Xing},
  \citenamefont {Wei},\ and\ \citenamefont {Liao}}]{xing2024quantum}%
  \BibitemOpen
  \bibfield  {author} {\bibinfo {author} {\bibfnamefont {F.}~\bibnamefont
  {Xing}}, \bibinfo {author} {\bibfnamefont {Y.}~\bibnamefont {Wei}},\ and\
  \bibinfo {author} {\bibfnamefont {Z.}~\bibnamefont {Liao}},\ }\bibfield
  {title} {\bibinfo {title} {Quantum search in many-body interacting systems
  with long-range interactions},\ }\href@noop {} {\bibfield  {journal}
  {\bibinfo  {journal} {Physical Review A}\ }\textbf {\bibinfo {volume}
  {109}},\ \bibinfo {pages} {052435} (\bibinfo {year} {2024})}\BibitemShut
  {NoStop}%
\bibitem [{\citenamefont {Wong}(2016)}]{wong2016spatial}%
  \BibitemOpen
  \bibfield  {author} {\bibinfo {author} {\bibfnamefont {T.~G.}\ \bibnamefont
  {Wong}},\ }\bibfield  {title} {\bibinfo {title} {Spatial search by
  continuous-time quantum walk with multiple marked vertices},\ }\href@noop {}
  {\bibfield  {journal} {\bibinfo  {journal} {Quantum Information Processing}\
  }\textbf {\bibinfo {volume} {15}},\ \bibinfo {pages} {1411} (\bibinfo {year}
  {2016})}\BibitemShut {NoStop}%
\bibitem [{\citenamefont {Wong}(2015)}]{wong2015grover}%
  \BibitemOpen
  \bibfield  {author} {\bibinfo {author} {\bibfnamefont {T.~G.}\ \bibnamefont
  {Wong}},\ }\bibfield  {title} {\bibinfo {title} {Grover search with
  lackadaisical quantum walks},\ }\href@noop {} {\bibfield  {journal} {\bibinfo
   {journal} {Journal of Physics A: Mathematical and Theoretical}\ }\textbf
  {\bibinfo {volume} {48}},\ \bibinfo {pages} {435304} (\bibinfo {year}
  {2015})}\BibitemShut {NoStop}%
\bibitem [{\citenamefont {Wang}\ \emph {et~al.}(2022)\citenamefont {Wang},
  \citenamefont {Qu}, \citenamefont {Wang}, \citenamefont {Chen},\ and\
  \citenamefont {Ma}}]{wang2022multiparticle}%
  \BibitemOpen
  \bibfield  {author} {\bibinfo {author} {\bibfnamefont {S.-M.}\ \bibnamefont
  {Wang}}, \bibinfo {author} {\bibfnamefont {Y.-J.}\ \bibnamefont {Qu}},
  \bibinfo {author} {\bibfnamefont {H.-W.}\ \bibnamefont {Wang}}, \bibinfo
  {author} {\bibfnamefont {Z.}~\bibnamefont {Chen}},\ and\ \bibinfo {author}
  {\bibfnamefont {H.-Y.}\ \bibnamefont {Ma}},\ }\bibfield  {title} {\bibinfo
  {title} {Multiparticle quantum walk--based error correction algorithm with
  two-lattice bose--hubbard model},\ }\href@noop {} {\bibfield  {journal}
  {\bibinfo  {journal} {Frontiers in Physics}\ }\textbf {\bibinfo {volume}
  {10}},\ \bibinfo {pages} {1016009} (\bibinfo {year} {2022})}\BibitemShut
  {NoStop}%
\bibitem [{\citenamefont {Childs}\ \emph {et~al.}(2013)\citenamefont {Childs},
  \citenamefont {Gosset},\ and\ \citenamefont {Webb}}]{childs2013universal}%
  \BibitemOpen
  \bibfield  {author} {\bibinfo {author} {\bibfnamefont {A.~M.}\ \bibnamefont
  {Childs}}, \bibinfo {author} {\bibfnamefont {D.}~\bibnamefont {Gosset}},\
  and\ \bibinfo {author} {\bibfnamefont {Z.}~\bibnamefont {Webb}},\ }\bibfield
  {title} {\bibinfo {title} {Universal computation by multiparticle quantum
  walk},\ }\href@noop {} {\bibfield  {journal} {\bibinfo  {journal} {Science}\
  }\textbf {\bibinfo {volume} {339}},\ \bibinfo {pages} {791} (\bibinfo {year}
  {2013})}\BibitemShut {NoStop}%
\bibitem [{\citenamefont {Underwood}\ and\ \citenamefont
  {Feder}(2012)}]{underwood2012bose}%
  \BibitemOpen
  \bibfield  {author} {\bibinfo {author} {\bibfnamefont {M.~S.}\ \bibnamefont
  {Underwood}}\ and\ \bibinfo {author} {\bibfnamefont {D.~L.}\ \bibnamefont
  {Feder}},\ }\bibfield  {title} {\bibinfo {title} {Bose-hubbard model for
  universal quantum-walk-based computation},\ }\href@noop {} {\bibfield
  {journal} {\bibinfo  {journal} {Physical Review A}\ }\textbf {\bibinfo
  {volume} {85}},\ \bibinfo {pages} {052314} (\bibinfo {year}
  {2012})}\BibitemShut {NoStop}%
\bibitem [{\citenamefont {e~Silva}\ and\ \citenamefont
  {Brod}(2024)}]{e2024two}%
  \BibitemOpen
  \bibfield  {author} {\bibinfo {author} {\bibfnamefont {L.~L.}\ \bibnamefont
  {e~Silva}}\ and\ \bibinfo {author} {\bibfnamefont {D.~J.}\ \bibnamefont
  {Brod}},\ }\bibfield  {title} {\bibinfo {title} {Two-particle scattering on
  non-translation invariant line lattices},\ }\href@noop {} {\bibfield
  {journal} {\bibinfo  {journal} {Quantum}\ }\textbf {\bibinfo {volume} {8}},\
  \bibinfo {pages} {1308} (\bibinfo {year} {2024})}\BibitemShut {NoStop}%
\bibitem [{\citenamefont {Asaka}\ \emph {et~al.}(2023)\citenamefont {Asaka},
  \citenamefont {Sakai},\ and\ \citenamefont {Yahagi}}]{asaka2023two}%
  \BibitemOpen
  \bibfield  {author} {\bibinfo {author} {\bibfnamefont {R.}~\bibnamefont
  {Asaka}}, \bibinfo {author} {\bibfnamefont {K.}~\bibnamefont {Sakai}},\ and\
  \bibinfo {author} {\bibfnamefont {R.}~\bibnamefont {Yahagi}},\ }\bibfield
  {title} {\bibinfo {title} {Two-level quantum walkers on directed graphs. i.
  universal quantum computing},\ }\href@noop {} {\bibfield  {journal} {\bibinfo
   {journal} {Physical Review A}\ }\textbf {\bibinfo {volume} {107}},\ \bibinfo
  {pages} {022415} (\bibinfo {year} {2023})}\BibitemShut {NoStop}%
\bibitem [{\citenamefont {Lahini}\ \emph {et~al.}(2018)\citenamefont {Lahini},
  \citenamefont {Steinbrecher}, \citenamefont {Bookatz},\ and\ \citenamefont
  {Englund}}]{lahini2018quantum}%
  \BibitemOpen
  \bibfield  {author} {\bibinfo {author} {\bibfnamefont {Y.}~\bibnamefont
  {Lahini}}, \bibinfo {author} {\bibfnamefont {G.~R.}\ \bibnamefont
  {Steinbrecher}}, \bibinfo {author} {\bibfnamefont {A.~D.}\ \bibnamefont
  {Bookatz}},\ and\ \bibinfo {author} {\bibfnamefont {D.}~\bibnamefont
  {Englund}},\ }\bibfield  {title} {\bibinfo {title} {Quantum logic using
  correlated one-dimensional quantum walks},\ }\href@noop {} {\bibfield
  {journal} {\bibinfo  {journal} {npj Quantum Information}\ }\textbf {\bibinfo
  {volume} {4}},\ \bibinfo {pages} {2} (\bibinfo {year} {2018})}\BibitemShut
  {NoStop}%
\bibitem [{\citenamefont {Koch}\ \emph {et~al.}(2007)\citenamefont {Koch},
  \citenamefont {Yu}, \citenamefont {Gambetta}, \citenamefont {Houck},
  \citenamefont {Schuster}, \citenamefont {Majer}, \citenamefont {Blais},
  \citenamefont {Devoret}, \citenamefont {Girvin},\ and\ \citenamefont
  {Schoelkopf}}]{koch2007charge}%
  \BibitemOpen
  \bibfield  {author} {\bibinfo {author} {\bibfnamefont {J.}~\bibnamefont
  {Koch}}, \bibinfo {author} {\bibfnamefont {T.~M.}\ \bibnamefont {Yu}},
  \bibinfo {author} {\bibfnamefont {J.}~\bibnamefont {Gambetta}}, \bibinfo
  {author} {\bibfnamefont {A.~A.}\ \bibnamefont {Houck}}, \bibinfo {author}
  {\bibfnamefont {D.~I.}\ \bibnamefont {Schuster}}, \bibinfo {author}
  {\bibfnamefont {J.}~\bibnamefont {Majer}}, \bibinfo {author} {\bibfnamefont
  {A.}~\bibnamefont {Blais}}, \bibinfo {author} {\bibfnamefont {M.~H.}\
  \bibnamefont {Devoret}}, \bibinfo {author} {\bibfnamefont {S.~M.}\
  \bibnamefont {Girvin}},\ and\ \bibinfo {author} {\bibfnamefont {R.~J.}\
  \bibnamefont {Schoelkopf}},\ }\bibfield  {title} {\bibinfo {title}
  {Charge-insensitive qubit design derived from the cooper pair box},\
  }\href@noop {} {\bibfield  {journal} {\bibinfo  {journal} {Physical Review
  A—Atomic, Molecular, and Optical Physics}\ }\textbf {\bibinfo {volume}
  {76}},\ \bibinfo {pages} {042319} (\bibinfo {year} {2007})}\BibitemShut
  {NoStop}%
\bibitem [{\citenamefont {Hutchings}\ \emph {et~al.}(2017)\citenamefont
  {Hutchings}, \citenamefont {Hertzberg}, \citenamefont {Liu}, \citenamefont
  {Bronn}, \citenamefont {Keefe}, \citenamefont {Brink}, \citenamefont {Chow},\
  and\ \citenamefont {Plourde}}]{hutchings2017tunable}%
  \BibitemOpen
  \bibfield  {author} {\bibinfo {author} {\bibfnamefont {M.}~\bibnamefont
  {Hutchings}}, \bibinfo {author} {\bibfnamefont {J.~B.}\ \bibnamefont
  {Hertzberg}}, \bibinfo {author} {\bibfnamefont {Y.}~\bibnamefont {Liu}},
  \bibinfo {author} {\bibfnamefont {N.~T.}\ \bibnamefont {Bronn}}, \bibinfo
  {author} {\bibfnamefont {G.~A.}\ \bibnamefont {Keefe}}, \bibinfo {author}
  {\bibfnamefont {M.}~\bibnamefont {Brink}}, \bibinfo {author} {\bibfnamefont
  {J.~M.}\ \bibnamefont {Chow}},\ and\ \bibinfo {author} {\bibfnamefont
  {B.}~\bibnamefont {Plourde}},\ }\bibfield  {title} {\bibinfo {title} {Tunable
  superconducting qubits with flux-independent coherence},\ }\href@noop {}
  {\bibfield  {journal} {\bibinfo  {journal} {Physical Review Applied}\
  }\textbf {\bibinfo {volume} {8}},\ \bibinfo {pages} {044003} (\bibinfo {year}
  {2017})}\BibitemShut {NoStop}%
\bibitem [{\citenamefont {Steffen}\ \emph {et~al.}(2010)\citenamefont
  {Steffen}, \citenamefont {Kumar}, \citenamefont {DiVincenzo}, \citenamefont
  {Rozen}, \citenamefont {Keefe}, \citenamefont {Rothwell},\ and\ \citenamefont
  {Ketchen}}]{steffen2010high}%
  \BibitemOpen
  \bibfield  {author} {\bibinfo {author} {\bibfnamefont {M.}~\bibnamefont
  {Steffen}}, \bibinfo {author} {\bibfnamefont {S.}~\bibnamefont {Kumar}},
  \bibinfo {author} {\bibfnamefont {D.~P.}\ \bibnamefont {DiVincenzo}},
  \bibinfo {author} {\bibfnamefont {J.~R.}\ \bibnamefont {Rozen}}, \bibinfo
  {author} {\bibfnamefont {G.~A.}\ \bibnamefont {Keefe}}, \bibinfo {author}
  {\bibfnamefont {.~f. M.~B.}\ \bibnamefont {Rothwell}},\ and\ \bibinfo
  {author} {\bibfnamefont {M.~B.}\ \bibnamefont {Ketchen}},\ }\bibfield
  {title} {\bibinfo {title} {High-coherence hybrid superconducting qubit},\
  }\href@noop {} {\bibfield  {journal} {\bibinfo  {journal} {Physical review
  letters}\ }\textbf {\bibinfo {volume} {105}},\ \bibinfo {pages} {100502}
  (\bibinfo {year} {2010})}\BibitemShut {NoStop}%
\bibitem [{\citenamefont {Yan}\ \emph {et~al.}(2016)\citenamefont {Yan},
  \citenamefont {Gustavsson}, \citenamefont {Kamal}, \citenamefont {Birenbaum},
  \citenamefont {Sears}, \citenamefont {Hover}, \citenamefont {Gudmundsen},
  \citenamefont {Rosenberg}, \citenamefont {Samach}, \citenamefont {Weber}
  \emph {et~al.}}]{yan2016flux}%
  \BibitemOpen
  \bibfield  {author} {\bibinfo {author} {\bibfnamefont {F.}~\bibnamefont
  {Yan}}, \bibinfo {author} {\bibfnamefont {S.}~\bibnamefont {Gustavsson}},
  \bibinfo {author} {\bibfnamefont {A.}~\bibnamefont {Kamal}}, \bibinfo
  {author} {\bibfnamefont {J.}~\bibnamefont {Birenbaum}}, \bibinfo {author}
  {\bibfnamefont {A.~P.}\ \bibnamefont {Sears}}, \bibinfo {author}
  {\bibfnamefont {D.}~\bibnamefont {Hover}}, \bibinfo {author} {\bibfnamefont
  {T.~J.}\ \bibnamefont {Gudmundsen}}, \bibinfo {author} {\bibfnamefont
  {D.}~\bibnamefont {Rosenberg}}, \bibinfo {author} {\bibfnamefont
  {G.}~\bibnamefont {Samach}}, \bibinfo {author} {\bibfnamefont
  {S.}~\bibnamefont {Weber}}, \emph {et~al.},\ }\bibfield  {title} {\bibinfo
  {title} {The flux qubit revisited to enhance coherence and reproducibility},\
  }\href@noop {} {\bibfield  {journal} {\bibinfo  {journal} {Nature
  communications}\ }\textbf {\bibinfo {volume} {7}},\ \bibinfo {pages} {12964}
  (\bibinfo {year} {2016})}\BibitemShut {NoStop}%
\bibitem [{\citenamefont {Yan}\ \emph {et~al.}(2018)\citenamefont {Yan},
  \citenamefont {Krantz}, \citenamefont {Sung}, \citenamefont {Kjaergaard},
  \citenamefont {Campbell}, \citenamefont {Orlando}, \citenamefont
  {Gustavsson},\ and\ \citenamefont {Oliver}}]{yan2018tunable}%
  \BibitemOpen
  \bibfield  {author} {\bibinfo {author} {\bibfnamefont {F.}~\bibnamefont
  {Yan}}, \bibinfo {author} {\bibfnamefont {P.}~\bibnamefont {Krantz}},
  \bibinfo {author} {\bibfnamefont {Y.}~\bibnamefont {Sung}}, \bibinfo {author}
  {\bibfnamefont {M.}~\bibnamefont {Kjaergaard}}, \bibinfo {author}
  {\bibfnamefont {D.~L.}\ \bibnamefont {Campbell}}, \bibinfo {author}
  {\bibfnamefont {T.~P.}\ \bibnamefont {Orlando}}, \bibinfo {author}
  {\bibfnamefont {S.}~\bibnamefont {Gustavsson}},\ and\ \bibinfo {author}
  {\bibfnamefont {W.~D.}\ \bibnamefont {Oliver}},\ }\bibfield  {title}
  {\bibinfo {title} {Tunable coupling scheme for implementing high-fidelity
  two-qubit gates},\ }\href@noop {} {\bibfield  {journal} {\bibinfo  {journal}
  {Physical Review Applied}\ }\textbf {\bibinfo {volume} {10}},\ \bibinfo
  {pages} {054062} (\bibinfo {year} {2018})}\BibitemShut {NoStop}%
\bibitem [{\citenamefont {Sung}\ \emph {et~al.}(2021)\citenamefont {Sung},
  \citenamefont {Ding}, \citenamefont {Braum{\"u}ller}, \citenamefont
  {Veps{\"a}l{\"a}inen}, \citenamefont {Kannan}, \citenamefont {Kjaergaard},
  \citenamefont {Greene}, \citenamefont {Samach}, \citenamefont {McNally},
  \citenamefont {Kim} \emph {et~al.}}]{sung2021realization}%
  \BibitemOpen
  \bibfield  {author} {\bibinfo {author} {\bibfnamefont {Y.}~\bibnamefont
  {Sung}}, \bibinfo {author} {\bibfnamefont {L.}~\bibnamefont {Ding}}, \bibinfo
  {author} {\bibfnamefont {J.}~\bibnamefont {Braum{\"u}ller}}, \bibinfo
  {author} {\bibfnamefont {A.}~\bibnamefont {Veps{\"a}l{\"a}inen}}, \bibinfo
  {author} {\bibfnamefont {B.}~\bibnamefont {Kannan}}, \bibinfo {author}
  {\bibfnamefont {M.}~\bibnamefont {Kjaergaard}}, \bibinfo {author}
  {\bibfnamefont {A.}~\bibnamefont {Greene}}, \bibinfo {author} {\bibfnamefont
  {G.~O.}\ \bibnamefont {Samach}}, \bibinfo {author} {\bibfnamefont
  {C.}~\bibnamefont {McNally}}, \bibinfo {author} {\bibfnamefont
  {D.}~\bibnamefont {Kim}}, \emph {et~al.},\ }\bibfield  {title} {\bibinfo
  {title} {Realization of high-fidelity cz and z z-free iswap gates with a
  tunable coupler},\ }\href@noop {} {\bibfield  {journal} {\bibinfo  {journal}
  {Physical Review X}\ }\textbf {\bibinfo {volume} {11}},\ \bibinfo {pages}
  {021058} (\bibinfo {year} {2021})}\BibitemShut {NoStop}%
\bibitem [{\citenamefont {Takayanagi}(2016)}]{takayanagi2016effective}%
  \BibitemOpen
  \bibfield  {author} {\bibinfo {author} {\bibfnamefont {K.}~\bibnamefont
  {Takayanagi}},\ }\bibfield  {title} {\bibinfo {title} {Effective interaction
  in unified perturbation theory},\ }\href@noop {} {\bibfield  {journal}
  {\bibinfo  {journal} {Annals of Physics}\ }\textbf {\bibinfo {volume}
  {364}},\ \bibinfo {pages} {200} (\bibinfo {year} {2016})}\BibitemShut
  {NoStop}%
\bibitem [{\citenamefont {Cederbaum}\ \emph {et~al.}(1989)\citenamefont
  {Cederbaum}, \citenamefont {Schirmer},\ and\ \citenamefont
  {Meyer}}]{cederbaum1989block}%
  \BibitemOpen
  \bibfield  {author} {\bibinfo {author} {\bibfnamefont {L.}~\bibnamefont
  {Cederbaum}}, \bibinfo {author} {\bibfnamefont {J.}~\bibnamefont
  {Schirmer}},\ and\ \bibinfo {author} {\bibfnamefont {H.-D.}\ \bibnamefont
  {Meyer}},\ }\bibfield  {title} {\bibinfo {title} {Block diagonalisation of
  hermitian matrices},\ }\href@noop {} {\bibfield  {journal} {\bibinfo
  {journal} {Journal of physics A: Mathematical and General}\ }\textbf
  {\bibinfo {volume} {22}},\ \bibinfo {pages} {2427} (\bibinfo {year}
  {1989})}\BibitemShut {NoStop}%
\bibitem [{\citenamefont {Dutta}\ \emph {et~al.}(2015)\citenamefont {Dutta},
  \citenamefont {Gajda}, \citenamefont {Hauke}, \citenamefont {Lewenstein},
  \citenamefont {L{\"u}hmann}, \citenamefont {Malomed}, \citenamefont
  {Sowi{\'n}ski},\ and\ \citenamefont {Zakrzewski}}]{dutta2015non}%
  \BibitemOpen
  \bibfield  {author} {\bibinfo {author} {\bibfnamefont {O.}~\bibnamefont
  {Dutta}}, \bibinfo {author} {\bibfnamefont {M.}~\bibnamefont {Gajda}},
  \bibinfo {author} {\bibfnamefont {P.}~\bibnamefont {Hauke}}, \bibinfo
  {author} {\bibfnamefont {M.}~\bibnamefont {Lewenstein}}, \bibinfo {author}
  {\bibfnamefont {D.-S.}\ \bibnamefont {L{\"u}hmann}}, \bibinfo {author}
  {\bibfnamefont {B.~A.}\ \bibnamefont {Malomed}}, \bibinfo {author}
  {\bibfnamefont {T.}~\bibnamefont {Sowi{\'n}ski}},\ and\ \bibinfo {author}
  {\bibfnamefont {J.}~\bibnamefont {Zakrzewski}},\ }\bibfield  {title}
  {\bibinfo {title} {Non-standard hubbard models in optical lattices: a
  review},\ }\href@noop {} {\bibfield  {journal} {\bibinfo  {journal} {Reports
  on Progress in Physics}\ }\textbf {\bibinfo {volume} {78}},\ \bibinfo {pages}
  {066001} (\bibinfo {year} {2015})}\BibitemShut {NoStop}%
\bibitem [{\citenamefont {Baier}\ \emph {et~al.}(2016)\citenamefont {Baier},
  \citenamefont {Mark}, \citenamefont {Petter}, \citenamefont {Aikawa},
  \citenamefont {Chomaz}, \citenamefont {Cai}, \citenamefont {Baranov},
  \citenamefont {Zoller},\ and\ \citenamefont {Ferlaino}}]{baier2016extended}%
  \BibitemOpen
  \bibfield  {author} {\bibinfo {author} {\bibfnamefont {S.}~\bibnamefont
  {Baier}}, \bibinfo {author} {\bibfnamefont {M.~J.}\ \bibnamefont {Mark}},
  \bibinfo {author} {\bibfnamefont {D.}~\bibnamefont {Petter}}, \bibinfo
  {author} {\bibfnamefont {K.}~\bibnamefont {Aikawa}}, \bibinfo {author}
  {\bibfnamefont {L.}~\bibnamefont {Chomaz}}, \bibinfo {author} {\bibfnamefont
  {Z.}~\bibnamefont {Cai}}, \bibinfo {author} {\bibfnamefont {M.}~\bibnamefont
  {Baranov}}, \bibinfo {author} {\bibfnamefont {P.}~\bibnamefont {Zoller}},\
  and\ \bibinfo {author} {\bibfnamefont {F.}~\bibnamefont {Ferlaino}},\
  }\bibfield  {title} {\bibinfo {title} {Extended bose-hubbard models with
  ultracold magnetic atoms},\ }\href@noop {} {\bibfield  {journal} {\bibinfo
  {journal} {Science}\ }\textbf {\bibinfo {volume} {352}},\ \bibinfo {pages}
  {201} (\bibinfo {year} {2016})}\BibitemShut {NoStop}%
\bibitem [{\citenamefont {Browaeys}\ and\ \citenamefont
  {Lahaye}(2020)}]{browaeys2020many}%
  \BibitemOpen
  \bibfield  {author} {\bibinfo {author} {\bibfnamefont {A.}~\bibnamefont
  {Browaeys}}\ and\ \bibinfo {author} {\bibfnamefont {T.}~\bibnamefont
  {Lahaye}},\ }\bibfield  {title} {\bibinfo {title} {Many-body physics with
  individually controlled rydberg atoms},\ }\href@noop {} {\bibfield  {journal}
  {\bibinfo  {journal} {Nature Physics}\ }\textbf {\bibinfo {volume} {16}},\
  \bibinfo {pages} {132} (\bibinfo {year} {2020})}\BibitemShut {NoStop}%
\bibitem [{\citenamefont {Ebadi}\ \emph {et~al.}(2022)\citenamefont {Ebadi},
  \citenamefont {Keesling}, \citenamefont {Cain}, \citenamefont {Wang},
  \citenamefont {Levine}, \citenamefont {Bluvstein}, \citenamefont {Semeghini},
  \citenamefont {Omran}, \citenamefont {Liu}, \citenamefont {Samajdar} \emph
  {et~al.}}]{ebadi2022quantum}%
  \BibitemOpen
  \bibfield  {author} {\bibinfo {author} {\bibfnamefont {S.}~\bibnamefont
  {Ebadi}}, \bibinfo {author} {\bibfnamefont {A.}~\bibnamefont {Keesling}},
  \bibinfo {author} {\bibfnamefont {M.}~\bibnamefont {Cain}}, \bibinfo {author}
  {\bibfnamefont {T.~T.}\ \bibnamefont {Wang}}, \bibinfo {author}
  {\bibfnamefont {H.}~\bibnamefont {Levine}}, \bibinfo {author} {\bibfnamefont
  {D.}~\bibnamefont {Bluvstein}}, \bibinfo {author} {\bibfnamefont
  {G.}~\bibnamefont {Semeghini}}, \bibinfo {author} {\bibfnamefont
  {A.}~\bibnamefont {Omran}}, \bibinfo {author} {\bibfnamefont {J.-G.}\
  \bibnamefont {Liu}}, \bibinfo {author} {\bibfnamefont {R.}~\bibnamefont
  {Samajdar}}, \emph {et~al.},\ }\bibfield  {title} {\bibinfo {title} {Quantum
  optimization of maximum independent set using rydberg atom arrays},\
  }\href@noop {} {\bibfield  {journal} {\bibinfo  {journal} {Science}\ }\textbf
  {\bibinfo {volume} {376}},\ \bibinfo {pages} {1209} (\bibinfo {year}
  {2022})}\BibitemShut {NoStop}%
\bibitem [{\citenamefont {Kounalakis}\ \emph {et~al.}(2018)\citenamefont
  {Kounalakis}, \citenamefont {Dickel}, \citenamefont {Bruno}, \citenamefont
  {Langford},\ and\ \citenamefont {Steele}}]{kounalakis2018tuneable}%
  \BibitemOpen
  \bibfield  {author} {\bibinfo {author} {\bibfnamefont {M.}~\bibnamefont
  {Kounalakis}}, \bibinfo {author} {\bibfnamefont {C.}~\bibnamefont {Dickel}},
  \bibinfo {author} {\bibfnamefont {A.}~\bibnamefont {Bruno}}, \bibinfo
  {author} {\bibfnamefont {N.}~\bibnamefont {Langford}},\ and\ \bibinfo
  {author} {\bibfnamefont {G.}~\bibnamefont {Steele}},\ }\bibfield  {title}
  {\bibinfo {title} {Tuneable hopping and nonlinear cross-kerr interactions in
  a high-coherence superconducting circuit},\ }\href@noop {} {\bibfield
  {journal} {\bibinfo  {journal} {npj Quantum Information}\ }\textbf {\bibinfo
  {volume} {4}},\ \bibinfo {pages} {38} (\bibinfo {year} {2018})}\BibitemShut
  {NoStop}%
\bibitem [{\citenamefont {Lagoin}\ \emph {et~al.}(2022)\citenamefont {Lagoin},
  \citenamefont {Bhattacharya}, \citenamefont {Grass}, \citenamefont
  {Chhajlany}, \citenamefont {Salamon}, \citenamefont {Baldwin}, \citenamefont
  {Pfeiffer}, \citenamefont {Lewenstein}, \citenamefont {Holzmann},\ and\
  \citenamefont {Dubin}}]{lagoin2022extended}%
  \BibitemOpen
  \bibfield  {author} {\bibinfo {author} {\bibfnamefont {C.}~\bibnamefont
  {Lagoin}}, \bibinfo {author} {\bibfnamefont {U.}~\bibnamefont
  {Bhattacharya}}, \bibinfo {author} {\bibfnamefont {T.}~\bibnamefont {Grass}},
  \bibinfo {author} {\bibfnamefont {R.}~\bibnamefont {Chhajlany}}, \bibinfo
  {author} {\bibfnamefont {T.}~\bibnamefont {Salamon}}, \bibinfo {author}
  {\bibfnamefont {K.}~\bibnamefont {Baldwin}}, \bibinfo {author} {\bibfnamefont
  {L.}~\bibnamefont {Pfeiffer}}, \bibinfo {author} {\bibfnamefont
  {M.}~\bibnamefont {Lewenstein}}, \bibinfo {author} {\bibfnamefont
  {M.}~\bibnamefont {Holzmann}},\ and\ \bibinfo {author} {\bibfnamefont
  {F.}~\bibnamefont {Dubin}},\ }\bibfield  {title} {\bibinfo {title} {Extended
  bose--hubbard model with dipolar excitons},\ }\href@noop {} {\bibfield
  {journal} {\bibinfo  {journal} {Nature}\ }\textbf {\bibinfo {volume} {609}},\
  \bibinfo {pages} {485} (\bibinfo {year} {2022})}\BibitemShut {NoStop}%
\bibitem [{\citenamefont {Benatti}\ \emph {et~al.}(2020)\citenamefont
  {Benatti}, \citenamefont {Floreanini}, \citenamefont {Franchini},\ and\
  \citenamefont {Marzolino}}]{benatti2020entanglement}%
  \BibitemOpen
  \bibfield  {author} {\bibinfo {author} {\bibfnamefont {F.}~\bibnamefont
  {Benatti}}, \bibinfo {author} {\bibfnamefont {R.}~\bibnamefont {Floreanini}},
  \bibinfo {author} {\bibfnamefont {F.}~\bibnamefont {Franchini}},\ and\
  \bibinfo {author} {\bibfnamefont {U.}~\bibnamefont {Marzolino}},\ }\bibfield
  {title} {\bibinfo {title} {Entanglement in indistinguishable particle
  systems},\ }\href@noop {} {\bibfield  {journal} {\bibinfo  {journal} {Physics
  Reports}\ }\textbf {\bibinfo {volume} {878}},\ \bibinfo {pages} {1} (\bibinfo
  {year} {2020})}\BibitemShut {NoStop}%
\bibitem [{\citenamefont {Nielsen}\ and\ \citenamefont
  {Chuang}(2001)}]{nielsen2001quantum}%
  \BibitemOpen
  \bibfield  {author} {\bibinfo {author} {\bibfnamefont {M.~A.}\ \bibnamefont
  {Nielsen}}\ and\ \bibinfo {author} {\bibfnamefont {I.~L.}\ \bibnamefont
  {Chuang}},\ }\href@noop {} {\emph {\bibinfo {title} {Quantum computation and
  quantum information}}},\ Vol.~\bibinfo {volume} {2}\ (\bibinfo  {publisher}
  {Cambridge university press Cambridge},\ \bibinfo {year} {2001})\BibitemShut
  {NoStop}%
\bibitem [{\citenamefont {DiVincenzo}(1995)}]{divincenzo1995two}%
  \BibitemOpen
  \bibfield  {author} {\bibinfo {author} {\bibfnamefont {D.~P.}\ \bibnamefont
  {DiVincenzo}},\ }\bibfield  {title} {\bibinfo {title} {Two-bit gates are
  universal for quantum computation},\ }\href@noop {} {\bibfield  {journal}
  {\bibinfo  {journal} {Physical Review A}\ }\textbf {\bibinfo {volume} {51}},\
  \bibinfo {pages} {1015} (\bibinfo {year} {1995})}\BibitemShut {NoStop}%
\bibitem [{\citenamefont {Mundada}\ \emph {et~al.}(2019)\citenamefont
  {Mundada}, \citenamefont {Zhang}, \citenamefont {Hazard},\ and\ \citenamefont
  {Houck}}]{mundada2019suppression}%
  \BibitemOpen
  \bibfield  {author} {\bibinfo {author} {\bibfnamefont {P.}~\bibnamefont
  {Mundada}}, \bibinfo {author} {\bibfnamefont {G.}~\bibnamefont {Zhang}},
  \bibinfo {author} {\bibfnamefont {T.}~\bibnamefont {Hazard}},\ and\ \bibinfo
  {author} {\bibfnamefont {A.}~\bibnamefont {Houck}},\ }\bibfield  {title}
  {\bibinfo {title} {Suppression of qubit crosstalk in a tunable coupling
  superconducting circuit},\ }\href@noop {} {\bibfield  {journal} {\bibinfo
  {journal} {Physical Review Applied}\ }\textbf {\bibinfo {volume} {12}},\
  \bibinfo {pages} {054023} (\bibinfo {year} {2019})}\BibitemShut {NoStop}%
\bibitem [{\citenamefont {Pedersen}\ \emph {et~al.}(2007)\citenamefont
  {Pedersen}, \citenamefont {M{\o}ller},\ and\ \citenamefont
  {M{\o}lmer}}]{pedersen2007fidelity}%
  \BibitemOpen
  \bibfield  {author} {\bibinfo {author} {\bibfnamefont {L.~H.}\ \bibnamefont
  {Pedersen}}, \bibinfo {author} {\bibfnamefont {N.~M.}\ \bibnamefont
  {M{\o}ller}},\ and\ \bibinfo {author} {\bibfnamefont {K.}~\bibnamefont
  {M{\o}lmer}},\ }\bibfield  {title} {\bibinfo {title} {Fidelity of quantum
  operations},\ }\href@noop {} {\bibfield  {journal} {\bibinfo  {journal}
  {Physics Letters A}\ }\textbf {\bibinfo {volume} {367}},\ \bibinfo {pages}
  {47} (\bibinfo {year} {2007})}\BibitemShut {NoStop}%
\end{thebibliography}
\end{document}